\documentclass[]{aa} 
\usepackage[]{natbib}
\usepackage{graphicx,color}
\usepackage{txfonts}
\usepackage[colorlinks=true, citecolor=blue]{hyperref}

\newcommand{\orcid}[1]{\href{https://orcid.org/#1}{\includegraphics[width=10pt]{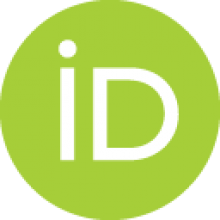}}}

\usepackage[normalem]{ulem}

\begin{document} 

\title{APOGEE-2S view of the globular cluster Patchick~125 (Gran~3)}
\subtitle{New metallicity and elemental abundances from high-resolution spectroscopy}

        \author{
                Jos\'e G. Fern\'andez-Trincado\inst{1}\thanks{E-mail:  jose.fernandez@ucn.cl and/or jfernandezt87@gmail.com}\orcid{0000-0003-3526-5052},
                Dante Minniti\inst{2,3}\orcid{0000-0002-7064-099X},
                Elisa R. Garro\inst{2}
       \and 
                Sandro Villanova\inst{4}\orcid{0000-0001-6205-1493}
}
        
        \authorrunning{J. G. Fern\'andez-Trincado et al.} 
        
\institute{
                Instituto de Astronom\'ia, Universidad Cat\'olica del Norte, Av. Angamos 0610, Antofagasta, Chile
                \and
                Depto. de Cs. F\'isicas, Facultad de Ciencias Exactas, Universidad Andr\'es Bello, Av. Fern\'andez Concha 700, Las Condes, Santiago, Chile
                \and 
                Vatican Observatory, V00120 Vatican City State, Italy
                \and
                Departamento de Astronom\'ia, Casilla 160-C, Universidad de Concepci\'on, Concepci\'on, Chile
          }
        
        \date{Received ...; Accepted ...}
        \titlerunning{XXXX}
        
        
        \abstract
        {We present detailed elemental abundances, radial velocity, and orbital elements for Patchick~125, a recently discovered metal-poor globular cluster (GC) in the direction of the Galactic bulge. Near-infrared high-resolution ($R\sim22,500$) spectra of two members were obtained during the second phase of the Apache Point Observatory Galactic Evolution Experiment at Las Campanas Observatory as part of the sixteenth Data Release (DR 16) of the Sloan Digital Sky Survey. We investigated elemental abundances for four chemical species, including $\alpha$- (Mg, Si), Fe-peak (Fe), and odd-Z (Al) elements. We find a metallicity covering the range from [Fe/H] $= -1.69$ to $-1.72$, suggesting that Patchick~125 likely exhibits a mean metallicity $\langle$[Fe/H]$\rangle \sim -1.7$, which represents a significant increase in metallicity for this cluster compared to previous low-resolution spectroscopic analyses. We also found a mean radial velocity of 95.9 km s$^{-1}$, which is $\sim$21.6 km s$^{-1}$ higher than reported in the literature. The observed stars  exhibit an $\alpha$-enrichment ([Mg/Fe]$\lesssim+0.20$, and [Si/Fe]$\lesssim +0.30$) that  follows the typical trend of metal-poor GCs. The aluminum abundance ratios for the present two member stars are enhanced in [Al/Fe]$\gtrsim +0.58$, which is a typical enrichment characteristic of  the so-called `second-generation' of stars in GCs at similar metallicity. This supports the possible presence of the multiple-population phenomenon in Patchick~125, as well as its genuine GC nature. Further, Patchick~125 shows a low-energy, low-eccentric ($<0.4$) and retrograde orbit captured by the inner Galaxy, near the edge of the bulge. We confirm that Patchick~125 is a genuine metal-poor GC, which is currently trapped in the vicinity of the Milky Way bulge.
                }
        
        \maketitle

        \section{Introduction}
        \label{section1}
        
Globular clusters (GCs) are among the most powerful cosmological archeology probes. Therefore, revealing in-depth details about the nature of these ancient star swarms can lead to a better understanding of the complex assembly scenarios that governed the early epochs of formation and evolution of their host galaxies. The bulge area of the Milky Way (MW) is plagued by a non-negligible fraction of GCs, many of which have remained hidden behind the high-absorption regions of the foreground field, and are affected by high levels of crowding and/or saturation by bright stars.

The combination of optical and near-infrared surveys, such as the Two-Micron All Sky Survey \citep[2MASS;][]{Skrutskie2006}, VISTA Variables in the V\'ia L\'actea \citep[VVV;][]{Minniti2010, Saito2012} and its extension \citep[VVVX;][]{Minniti2018a},  the ESA \textit{Gaia} mission \citep{Gaia2016, Gaia2018, Gaia2021} with its unprecedented astrometric precision, and the second generation of the Apache Point Observatory Galactic Evolution Experiment survey \citep[APOGEE-2;][]{Majewski2017}, which provides accurate spectroscopic and kinematics information, has allowed the extraordinary power of these ancient systems to be fully exploited, even in densely populated regions  and those heavily obscured by the interstellar medium (ISM).

Such is the case of  VVV~CL001 with E(B$-$V)$\sim2.2$, which was recently re-classified as the most metal-poor GC found so far ---with [Fe/H]$\sim -2.45$--- to have survived near the bulge region \citep{Fernandez-Trincado2021}, and NGC~6330 (or Tonantzintla~1) a highly reddened, E(B$-$V)$\sim$1.07 bulge GC, recently identified as the first known case of a relatively high-metallicity GC exhibiting evidence of correlation between its light- and heavy-elements and the presence of the phenomenon of multiple stellar populations \citep{Trincado2021}. In addition, UKS~1, a heavily reddened GC, with E(B$-$V)$\sim$2.62, was recently classified as a fossil relic of the bulge based on its chemodynamics properties \citep{Fernandez-Trincado2020}. These are examples of the wide gamut of highly reddened GCs examined so far towards the bulge area \citep[see][for other cases]{Saracino2015, Schiavon2017, Barbuy2018b, Fernandez-Trincado2019, Fernandez-Trincad2021b, Fernandez-Trincado2021c, Barbuy2018, Barbuy2021, Kunder2020, Kunder2021, Geisler2021, 2021TrincadoCapos, Alonso-Garcia2021}.

More recently, the VVV/VVVX survey has revealed that the census of Galactic GCs is still not complete, and more than 300 new low-luminosity GC candidates have been identified in the VVV/VVVX bulge$+$disc area toward the inner galaxy \citep[see e.g.][]{Minniti2020b, Minniti2021}, bulge region \citep{Minniti2017c, Minniti2017d, Minniti2018c, Minniti2018b, Camargo2019, Palma2019, Garro2021a, Obasi2021}, disc region \citep{Garro2020}, the extension of the Hrid halo stream \citep{Minniti2021b}, and the Sagittarius system \citep{Minniti2021c, Garro2021b}. A few of  those candidates have been confirmed as true GCs \citep[][]{Contreras2018, Gran2019, Barba2019, Villanova2019, Gran2021a, Romero-Colmenares2021, Dias2021}.

In this paper we present near-infrared (NIR) elemental abundances of the heavily obscured \citep[E(B$-$V)$\sim$1.06;][]{Garro2021c} GC Patchick~125 (originally discovered by Dana Patchick in 2016, internal communication) toward the Galactic bulge, which is positioned within  10.2\arcsec\ of the centre of the GC Gran~3 \citep{Gran2021b}, so both clusters are the same, and from here on we refer to this cluster as Patchick~125.

\begin{figure}
        \begin{center}
                \includegraphics[width=95mm]{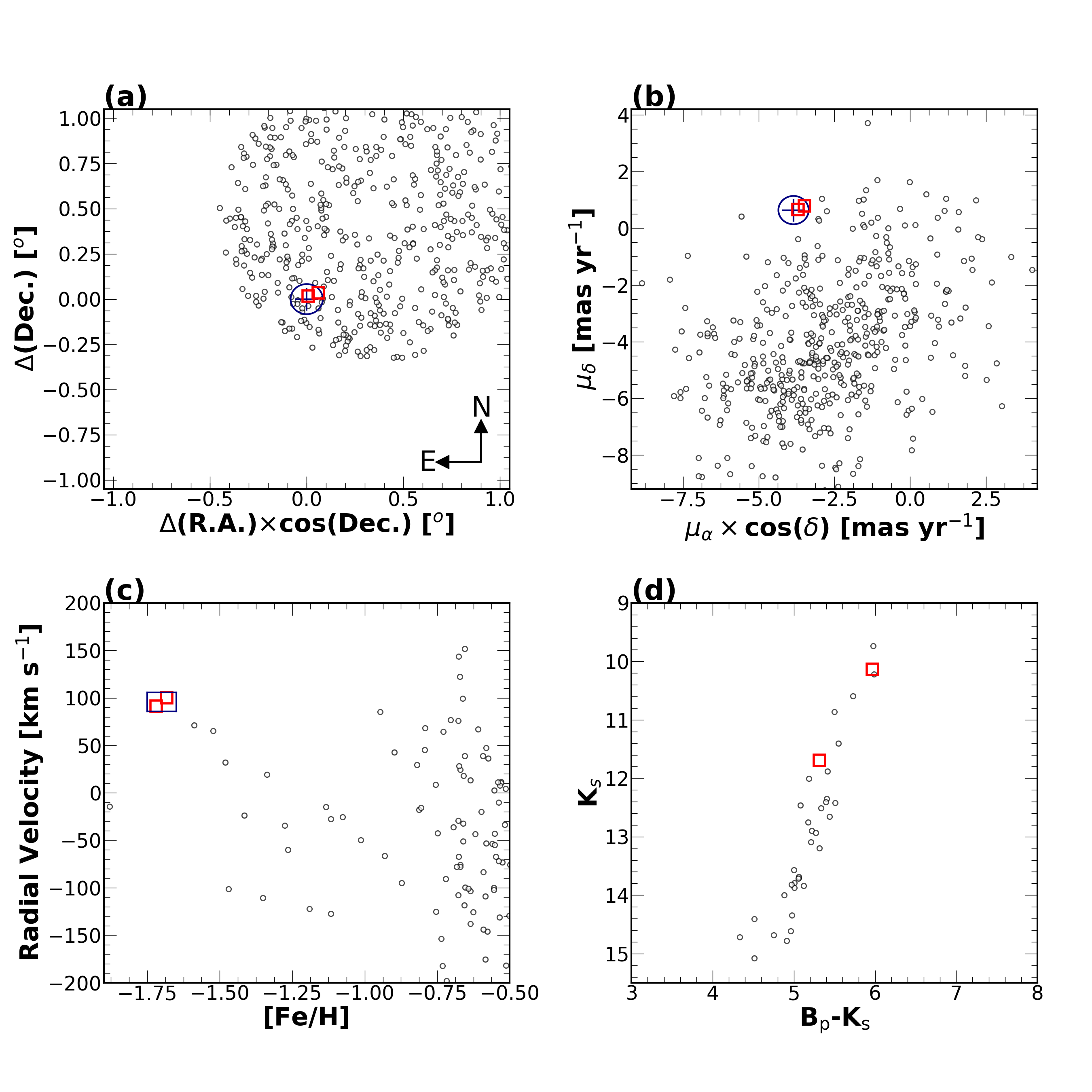}
                \caption{{\bf Main physical properties of Patchick~125 stars:} Panel (a) shows the spatial distribution of APOGEE-2S stars (empty circles). The position of Patchick~125 stars is shown with a navy cross, while the two cluster members analysed in this work are highlighted with red unfilled squares. A navy empty circle with a 5\arcmin radius and centred on Patchick~125 is shown for visual aid. Panel (b) shows the \textit{Gaia} eDR3 proper motion distribution of our sample. Panel (c) highlights the metallicity ([Fe/H]) versus radial velocity of APOGEE-2S stars with \texttt{ASPCAP}/DR16 determinations towards the Patchick~125 field (black circles) together with our \texttt{BACCHUS}-[Fe/H] determinations (red symbols). The blue box limited by $\pm 0.05$ dex and $\pm10$ km s$^{-1}$ and centred at [Fe/H]$= -1.7$ and radial velocity $=95.9$ km s$^{-1}$ encloses our potential cluster members. Panel (d) reveals the 2MASS$+$\textit{Gaia} eDR3 colour magnitude diagram corrected for differential reddening for Patchick stars within 3\arcmin.}
                \label{Figure1}
        \end{center}
\end{figure}            

\section{Spectroscopic data}

We searched for high-resolution ($R\sim22,500$) \textit{H}-band (1.51 --1.7 $\mu$m) spectra towards the field of the GC Patchick~125 ---at $\alpha =$ 17:05:00.70 and $\delta =$ $-$35:29:41.0--- in the publicly available sixteenth data release \citep[DR16;][]{Ahumada2020} of the APOGEE-2 survey \citep[][]{Majewski2017}, one of the programs within the Sloan Digital Sky Survey \citep{Blanton2017}. 

Stars in this field were observed using the APOGEE-2  spectrograph twin  \citep{Wilson2019} installed on the Ir\'en\'ee du Pont 2.5m telescope \citep{Bowen1973} at Las Campanas Observatory (APOGEE-2S). We refer the reader to \citet{Nidever2015}, \citet{Zamora2015}, \citet{Holtzman2015}, \citet{Garcia2016}, \citet{Zasowski2017}, \citet{Smith2021}, and \citet{Santana2021} for further details regarding the targeting strategy of the APOGEE-2S survey, spectra reduction, and analysis using the APOGEE Stellar Parameters and Chemical Abundance Pipeline (\texttt{ASPCAP}), the libraries of synthetic spectra, and the \textit{H}-band line list, respectively.

The APOGEE-2S plug-plate containing the Patchick~125 field was centred on ($l$, $b$) $\sim$ (350$^{\circ}$, $+$04$^{\circ}$) as part of the bulge program survey containing 493 science fibres. From these stars, we identified two potential sources within 5\arcmin\ of the centre of Patchick~125, which exhibit \textit{Gaia} eDR3 proper motions within a 0.5 mas yr$^{-1}$ radius around the mean proper motion of the cluster, $\mu_{\alpha} = -3.85\pm0.50$ and $\mu_{\delta}\cos(\delta) = +0.64\pm0.39$ \citep{Garro2021c}, which is in excellent agreement with the values reported in \citet{Gran2021b} of namely $\mu_{\alpha} = -3.78$ and $\mu_{\delta}\cos(\delta) = +0.66$. The APOGEE-2S radial velocities of these two stars also exhibit similar  kinematics, and lie in the region of the red giant branch (RGB) of Patchick~125. Figure \ref{Figure1} shows the main physical properties of the two newly identified members of Patchick~125 in the APOGEE-2S, \textit{Gaia} eDR3, and 2MASS footprint.

\begin{figure}
        \begin{center}
                \includegraphics[width=90mm]{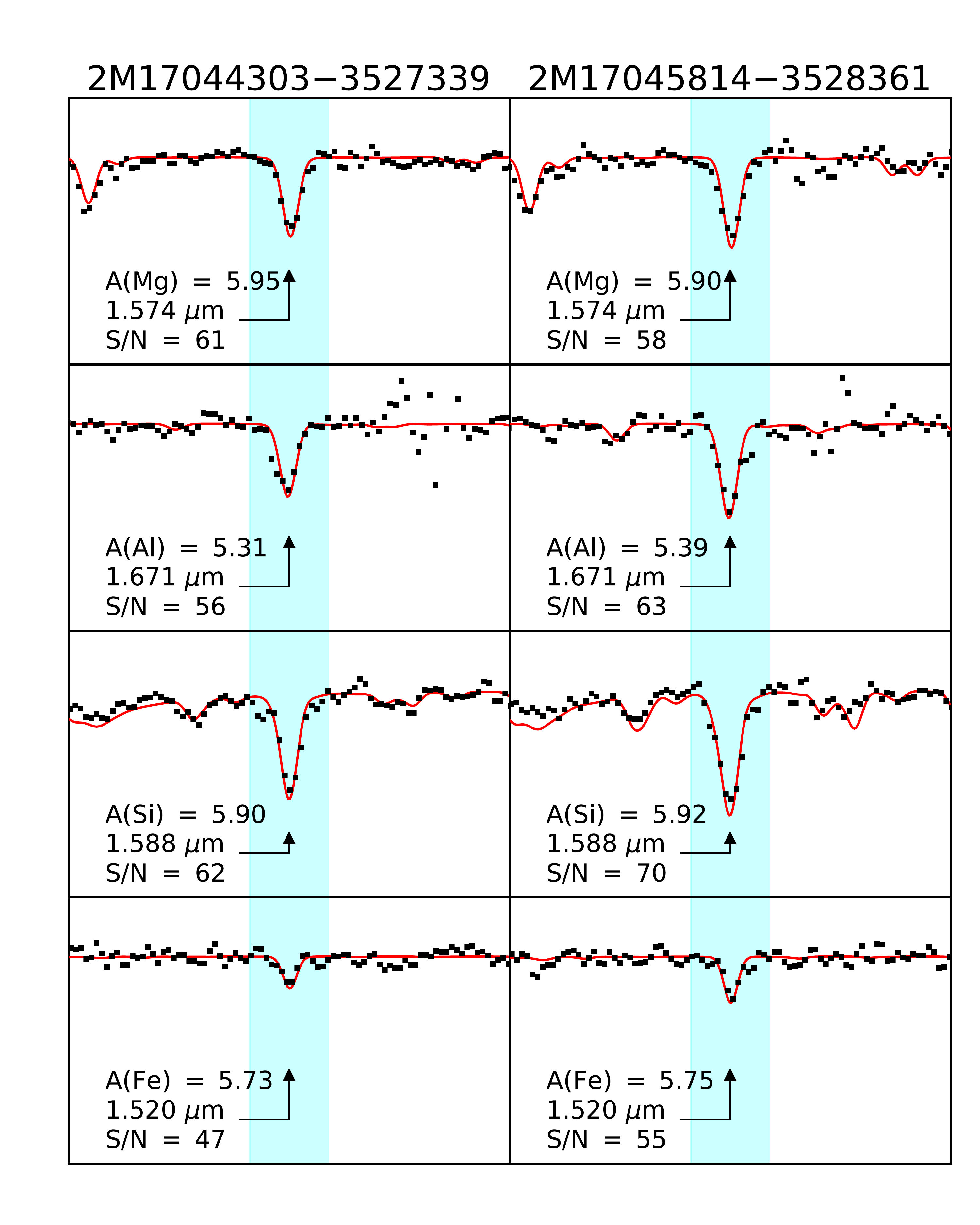}
                \caption{Discovery APOGEE-2S spectra of Patchick~125 members. Examples of selected Mg I, Al I, Si I, and Fe I lines are shown. Each panel shows the best-fit synthesis (red lines) from \texttt{BACCHUS} compared to the observed spectra (black symbols) of selected lines (marked with black arrows, and cyan shadow bands of 3.2$\times$10$^{-4}$ $\mu$m wide). }
                \label{Figure2}
        \end{center}
\end{figure}            

\section{Atmospheric parameters}
\label{atmospheric}

The atmospheric parameters for the two target stars were adopted in the same manner as described in \citet{Trincado2021}, that is, we applied a simple approach of fixing T$_{\rm eff}$ and $\log$ \textit{g} to values determined from optical$+$NIR photometric bands corrected for differential reddening. 

In summary, the colour-magnitude diagram (CMD) presented in Figure \ref{Figure1} was corrected for differential reddening using giant stars, and adopting the reddening law of \citet{Car89} and \citet{Donnell1994} and a total-to-selective absorption ratio RV$=$ 3.1. For this purpose, we selected all RGB stars within a radius of 3 arcmin around the cluster centre that have proper motions compatible with that of Patchick~125. First, we draw a ridge line along the RGB, and for each of the selected RGB stars we calculated its distance from this line along the reddening vector. The vertical projection of this distance gives the differential optical$+$NIR absorption at the position of the star, while the horizontal projection gives the differential optical$+$NIR reddening at the position of the star. After this first step, for each star of the field we selected the three nearest RGB stars, calculated the mean reddening and absorption, and finally subtracted these mean values from its optical$+$NIR colours and magnitudes. We underline the fact that the number of reference stars used for the reddening correction is the result of a compromise allowing a correction that is  affected as little as possible by photometric random errors whilst achieving the highest possible spatial resolution.

We obtain T$_{\rm eff}$ and $\log$ from photometry by determining the differential reddening-corrected CMD of Figure \ref{Figure1}(d). We then horizontally projected the position of each observed star until it intersected the \texttt{PARSEC} \citep{Bressan2012} isochrone (chosen to have an age of 12 Gyr, a metallicity of $-1.70,$ and an $\alpha$-enhancement of 0.3), and assumed T$_{\rm eff}$ and $\log$ \textit{g} to be the temperature and gravity at the point of the isochrones that has the same K$_{\rm s}$ magnitude as the star. We then applied a distance of 11.2 kpc and a reddening of E(B-V)$=$1.00 to the isochrone in order to obtain the best fit to the K$_{\rm s}$ versus B$_{\rm p}-$K$_{\rm s}$ CMD. We underline the fact that, for highly reddened objects like Patchick~125, the absorption corrections depend on their temperature. For this reason, we applied a temperature-dependent absorption correction to the isochrone. Without this, it is not possible to obtain a proper fit of the RGB, especially of the upper and cooler part.

\section{Elemental abundances}

The APOGEE-2S spectra of the newly identified GC Patchick~125 stars have a signal-to-noise (S/N) which is on the order of between 49 and 52 pixel$^{-1}$. This S/N level makes it difficult to obtain reliable elemental abundances for some chemical species commonly accessible from the \textit{H}-band of the APOGEE-2S survey. In particular, the signal from $^{12}$C$^{14}$N, $^{12}$C$^{16}$O, and $^{16}$OH molecules is too weak to provide reliable determinations of nitrogen, oxygen, and carbon. A visual examination of the whole spectra reveals that only four atomic elements can be accurately determined from the strengths of Mg I, Al I, Si I, and Fe I lines, as shown in Figure \ref{Figure2}, while the lines of other atomic elements are weak, and heavily affected by telluric features.

We made use of the Brussels Automatic Stellar Parameter code \citep[\texttt{BACCHUS;}][]{Masseron2016} to determine the elemental abundances of [Mg/Fe], [Al/Fe], [Si/Fe], and [Fe/H] in Patchick~125. Chemical abundances were derived from a local thermodynamics equilibrium (LTE) analysis using the \texttt{BACCHUS} combined with the \texttt{MARCS} model atmospheres \citep{Gustafsson2008}, and following the same technique as described in \citet{Fernandez-Trincad2021b}, and summarised here for guidance. With the atmospheric parameters determined in Section \ref{atmospheric}, the first step consisted in determining the metallicity from selected Fe I lines, the micro-turbulence velocity ($\xi_{t}$), and the convolution parameter. 

With the metallicity and main atmospheric parameters fixed, we then computed the abundance of each chemical species as follows: (a) We performed a synthesis using the full set of atomic line lists (Mg I, Al I, and Si I) described in \citet{Smith2021}. This set of lines is internally labelled as \texttt{linelist.20170418} based on the date of creation in the format YYYYMMDD. This was used to find the local continuum level via a linear fit. (b) We then performed cosmic ray and telluric line rejections, before (c) estimating the local S/N. (d) We automatically selected a series of flux points contributing to a given absorption line, and then (e) derived abundances by comparing the observed spectrum with a set of convolved synthetic spectra characterised by different abundances. Subsequently, four different abundance determination methods were used: (1) line-profile fitting; (2) core line-intensity comparison; (3) global goodness-of-fit estimate; and (4) an equivalent-width comparison. Each diagnostic yields validation flags. Based on these flags, a decision tree then rejects or accepts each estimate, keeping the best-fit abundance. We adopted the $\xi^{2}$ diagnostic as the abundance because of its robustness. However, we stored the information from the other diagnostics, including the standard deviation between all four methods. 

Table \ref{Table1} lists the final atmospheric parameters and the measured elemental abundances, while Figure \ref{Figure2} shows the line-by-line spectrum synthesis with the \texttt{BACCHUS} code around selected atomic Mg I, Al I, Si I, and Fe I lines. The uncertainties reported in Table \ref{Table1} were determined in the same manner as described in \citet{Fernandez-Trincado2019} by varying the atmospheric parameters one at a time by  $\pm100$ K for the effective temperature, $\pm 0.30$ cgs for the surface gravity, and $\pm 0.05$ km s$^{-1}$ for the microturbulent velocity, which are typical but conservative values. Thus, the reported uncertainties are defined as $\sigma^{2}_{total}  = \sigma^2_{T_{\rm eff}}    + \sigma^2_{{\rm log} g} + \sigma^2_{\xi_t}  + \sigma^2_{mean}$.

\begin{table}
                \begin{small}
        \begin{center}
                \setlength{\tabcolsep}{0.5mm}  
                \caption{APOGEE-2S elemental abundances of Patchick~125.}
                \begin{tabular}{|c|c|c|}
                        \hline
                         &  2M17044303$-$3527339 &  2M17045814$-$3528361  \\
                        \hline
                        T$_{\rm eff}$ (K)           &     $4631\pm100$            &  $4288\pm100$             \\
                        \hline
                        log \textit{g} (cgs)         &        $1.33\pm0.30$          &  $0.68\pm0.30$  \\
                        \hline
                        $\xi_{t}$ (km s$^{-1}$)      &          $1.87\pm0.05$  & $2.05\pm0.05$    \\
                        \hline
                        S/N (pixel$^{-1}$) &  52 & 49 \\
                        \hline
                        [Mg/Fe]    &       $+0.20\pm0.06$          &      $+0.13\pm0.07$     \\
                        \hline
                     [Al/Fe]   &       $+0.58\pm0.08$          &      $+0.83\pm0.09$      \\
           \hline                    
                        [Si/Fe]    &        $+0.28\pm0.08$         &       $+0.30\pm0.10$      \\
                        \hline
           [Fe/H]       &  $-1.72\pm0.09$          &        $-1.69\pm0.11$  \\
         \hline                         
                \end{tabular}  \label{Table1}
        \end{center}
                \end{small}
\end{table}     

\section{Abundance analysis}

We find that stars in Patchick~125 are as metal poor as $-1.69$ and $-1.72$, suggesting a mean metallicity of $\langle$[Fe/H]$\rangle\sim-1.70,$ which is consistent with recent photometric estimates for this cluster, [Fe/H]$= -1.8 \pm 0.2$ \citep{Garro2021c}. With a sample size of two stars, we cannot comment on the existence of a metallicity spread in Patchick~125. However, we plan to investigate the cluster's metallicity distribution in a future spectroscopic follow-up study.

It is important to note that \citet{Gran2021b} determined a mean metallicity of $\langle$[Fe/H]$\rangle = -2.33$ for Patchick~125 (or Gran~3), which is $\sim$0.63 more metal poor than our determinations. However, their methodology, which relies on the relation between CaT EW and the magnitude of the HB in the Johnson V filter, tends to be less precise than our high-resolution determinations which rely directly on the Fe I lines, as this cluster lies in a very highly reddened region, with E(B$-$V)$> 1$, and extinction ${\rm A_{v}} \sim 2.37$ \citep{Gran2021b}.

\begin{figure*}
        \begin{center}
                \includegraphics[width=180mm]{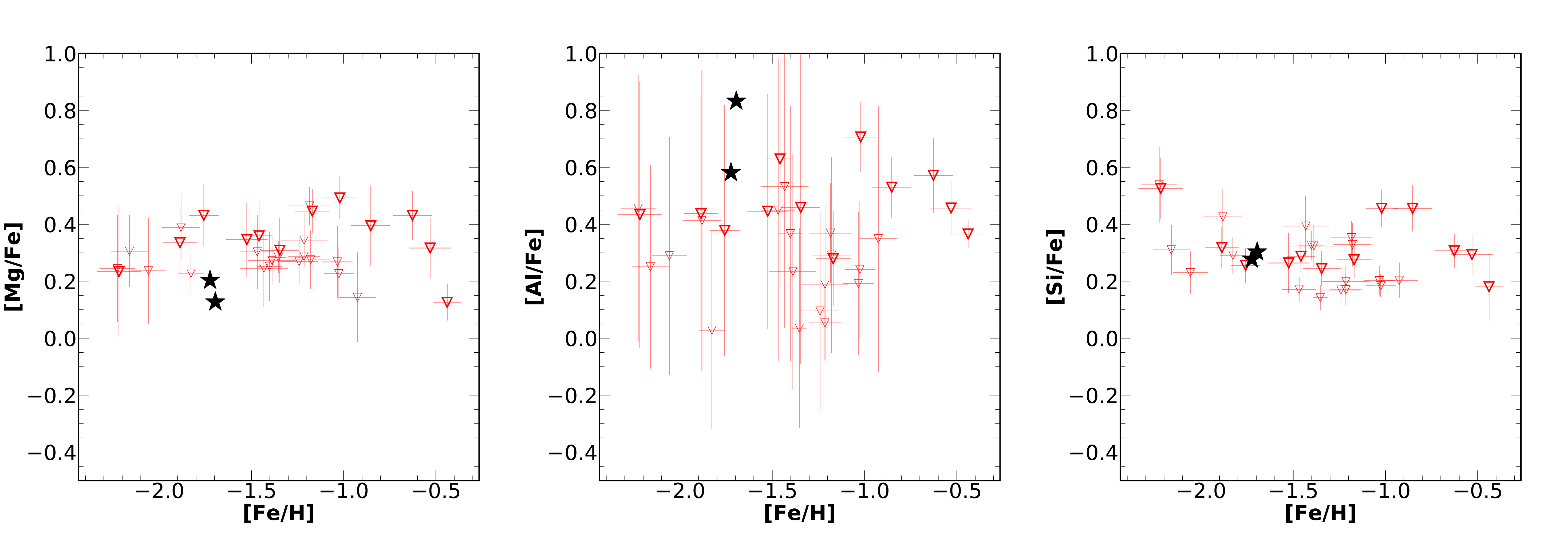}
                \caption{\texttt{BACCHUS} distribution of [Fe/H] vs. [Mg/Fe] (left), [Al/Fe] (middle), and [Si/Fe] (right) for GCs (red open symbols) analysed in \citet{Meszaros2020}. Patchick~125 members are highlighted with black star symbols.}
                \label{Figure3}
        \end{center}
\end{figure*}

The odd-Z element Al was found to be enhanced in the two Patchick~125  stars, placing this cluster well above the typical [Al/Fe] levels seen in other GCs, as can be seen in the middle panel in Figure \ref{Figure3}. The high [Al/Fe] abundance ratios measured in Patchick~125 could belong to the so-called second-generation population, which could explain the apparent overabundance in Al compared to other GCs.  Therefore, with the current limited sample in Patchick~125, it is not possible to reach a firm conclusion about the origin of this cluster based on the observed [Al/Fe]. 

The $\alpha$-element Mg is slightly overabundant compared to the Sun. Figure \ref{Figure3} shows that these two stars exhibit slightly lower [Mg/Fe] abundance ratios compared to those of metal-poor GCs, which could be due to the fact that these stars belong to the `second-generation' population with a high-aluminum enrichment, and therefore do not reflect the mean [Mg/Fe] enrichment of the cluster. Unfortunately, in this case, the [Mg/Fe] abundances alone are not useful indicators with which to distinguish between an in situ and accreted origin for this cluster.
        
The second measured $\alpha$-element Si shows that Patchick~125  stars exhibit [Si/Fe] abundance ratios comparable to those of  metal-poor GCs, as shown in the right panel of Figure \ref{Figure3}.
It is important to note that the high  [Al/Fe] enrichment well above $+0.58$ observed in Patchick~125 stars is typical among the so-called second-generation stars in GCs at similar metallicity \citep[M~55;][for instance]{Meszaros2020}, and is unlikely to be seen in dwarf galaxy populations \citep{Shetrone2001, Hasselquist2017, Hasselquist2021}, thus supporting the genuine GC nature of Patchick 125.

\section{Dynamical properties of Patchick~125}

We made use of the \texttt{GravPot16} model\footnote{\url{https://gravpot.utinam.cnrs.fr}} to predict the ensemble of orbits associated with Patchick~125. 

\subsection{Model}

We adopted the same \texttt{GravPot16} model configuration as described in \citet{Romero-Colmenares2021}, which consists of a boxy/peanut bar structure in the bulge region along with other composite stellar components belonging to the thin and thick discs, ISM, an oblate Hernquist stellar halo, and a dark-matter component characterised by an isothermal sphere truncated at $R_{gal} \sim 100$ kpc. For a more detailed description of the functional forms of the gravitational potential of each component, we refer the readers to a forthcoming paper (\texttt{GravPot16}; Fern\'andez-Trincado et al., in prep.). 

The structural parameters of our bar model (mass, present-day orientation, and pattern speed) are 1.1$\times$10$^{10}$ M$_{\odot}$, 20$^{\circ}$, and $\Omega_{\rm bar} = $41 km s$^{-1}$ kpc \citep{Sanders2019}, respectively, consistent with observational estimates. The bar scale lengths are $x_{0} =  1.46$ kpc, $y_{0} = 0.49$ kpc, and $z_{0} = 0.39$ kpc, where the effective boundary (or cut-off radius) of the bar on the \textit{x}-axis has a semi-major axis of 3.43 kpc.

For reference, the Galactic convention adopted in this work is: x-axis oriented towards $l = 0^{\circ}$ and $b = 0^{\circ}$, y-axis oriented towards $l = 90^{\circ}$ and $b = 0^{\circ}$, and the disc rotates towards $l = 90^{\circ}$; the velocity is also oriented along these directions. Following this convention, the Sun's orbital velocity vectors are [$U_{\odot}$,$V_{\odot}$,$W_{\odot}$] $=$ [11.1, 12.24, 7.25] km $s^{-1}$ \citep{Brunthaler2011}. The model has been rescaled to the Sun's Galactocentric distance of 8.3 kpc, and the local rotation velocity of 244.5 km s$^{-1}$ \citep{Sofue2015}.
\subsection{Orbit}

To compute the orbits of  Patchick~125 we adopted the following three initial conditions: (\textit{i}) The mean proper motions for Patchick~125 were adopted from the values determined by \citet{Garro2021c}, that is, $\mu_{\alpha} = -3.85\pm0.50$ and $\mu_{\delta}\cos(\delta) = +0.64\pm0.39$; (\textit{ii}) \citet{Garro2021c} measured distances of $D= 11.2$ kpc and $10.9$ kpc to this GC based on the optical and NIR CMDs, respectively. They also identified two RR Lyrae variable stars as cluster members based on their matching positions, magnitudes, and \textit{Gaia} proper motions. These are RR Lyrae-type ab stars \textit{Gaia} DR2 5977224553266268928 and 5977223144516980608 from \citet{Clementini2019}, both located at about 80$\arcsec$ from the cluster centre. We are able to check the determination of the distance for this cluster using the NIR photometry of these RRab stars; their magnitudes are $Ks= 14.845\pm0.025$  and $14.775 \pm 0.024$ mag, their colours are J$-$K$_{\rm s}=0.785 \pm 0.03$ and $0.589 \pm 0.03$ mag, and their periods are P$ = 0.601940$ and P$ = 0.738296$ days, respectively. Using the latest NIR PLZ relations of \citet{Bhardwaj2021} for [Fe/H]$=-1.7$, and adopting a field extinction of A$_{\rm K} = 0.327$ mag from \citet{Schlafly2011}, their distances are $D= 10.66 \pm 0.30$ kpc and $11.42 \pm 0.30$ kpc, respectively. This yields a mean cluster distance of $D= 11.0 \pm 0.5$ kpc, which is in agreement with the values obtained by \citet{Garro2021c}, and our (see Section \ref{atmospheric}). (\textit{iii}) The cluster mean radial velocity was assumed to be 95.9 km s$^{-1}$, which was obtained from our APOGEE-2S data. This adopted radial velocity value is in reasonable agreement with the value reported by \citep{Gran2021b}, of namely $74.23\pm2.70$ km s$^{-1}$, which is  lower than our value by $\sim$21.6 km s$^{-1}$. This difference in radial velocity could be due to systematic errors between the low resolution of the MUSE spectra and the high resolution of the APOGEE-2S spectra. However, we note that by adopting both values in radial velocity, our conclusions regarding the dynamical behaviour of Patchick~125 are not strongly affected. For our orbit computations, we assumed an error for the cluster radial velocity of the order of 10 km s$^{-1}$. 

We computed an ensemble of orbits by adopting a simple Monte Carlo approach and the Runge-Kutta algorithm of seventh to eighth order elaborated by \citet{F68}. The uncertainties in the input data were randomly propagated as 1$\sigma$ variation in a Gaussian Monte Carlo resampling. Thus, we ran 10 000 orbits, computed backwards in time for 1.5 Gyr. The 50$^{\rm th}$ percentile of the orbital elements were found for these 10 000 realisations, with uncertainty ranges given by the 16$^{\rm th}$ and 84$^{\rm th}$ percentiles. 

\begin{figure*}
        \begin{center}
                \includegraphics[width=180mm]{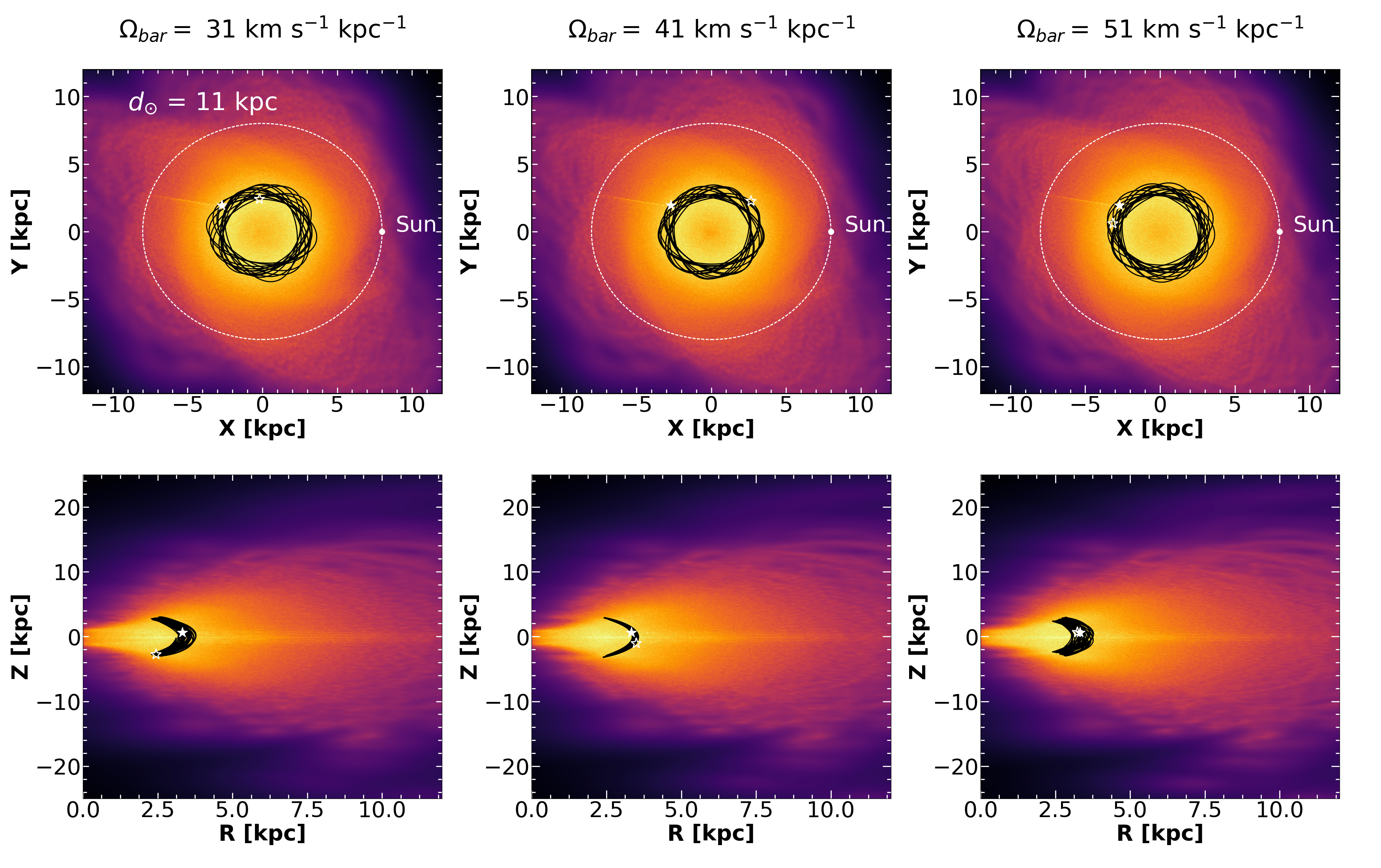}
                \caption{Ensemble of 10 000 orbits of Patchick~125 by considering the errors on the observables, projected on the equatorial and meridional Galactic planes in the inertial reference frame, with a bar pattern speed of 31, 41, and 51 km s$^{-1}$and  integrated over the past 1.5 Gyr. The red and orange colours correspond to more probable regions of the space, which are crossed most frequently by the simulated orbits. The black line shows the orbit of Patchick~125 from the observables without error bars. The white dashed line indicates the Sun's radius. The white filled and unfilled star symbols indicate the initial and final positions of the cluster, respectively.}
                \label{Figure4}
        \end{center}
\end{figure*}

The resulting orbital elements are listed in Table \ref{Table2} for three different values of  $\Omega_{\rm bar}$, which was varied in steps of 10 km s$^{-1}$ kpc in order to check for any significant impact of variations of this parameter. The minimal and maximum values of the z-component of the angular momentum in the inertial frame are also listed in this table, because this quantity is not conserved in a model with non-axisymmetric structures like \texttt{GravPot16}. Table \ref{Table2} and Figure \ref{Figure4} also show that our orbit results are not strongly affected by variations in $\Omega_{\rm bar}$.  

Figure \ref{Figure4} shows the probability densities of the resulting ensemble of orbits projected on the equatorial and meridional Galactic planes in the inertial reference frame. The red and yellow  colours correspond to more probable regions of space, which are crossed more frequently by the simulated orbits. 

Patchick~125 is found to have a low eccentricity and retrograde orbital configuration (see Table \ref{Table2}), which has perigalactocentric distances with incursions inside the cut-off radius of the bar, apogalactocentric distances (r$_{\rm apo} \sim $3.98 -- 4.51 kpc) at the edge of the MW bulge \citep[r$_{\rm edge}\sim3$ kpc;][]{Barbuy2018b, Perez-Villegas2020}, and vertical excursions from the Galactic plane, |Z|$_{\rm max} \lesssim 3.2$ kpc, which are similar to those recently found by \citet{Gran2021b}. Our observations caught Patchick~125 near the apocentre of its orbit.

\begin{table}
                \begin{small}
        \begin{center}
                \setlength{\tabcolsep}{1.0mm}  
                \caption{Orbital elements for Patchick~125.}
                \begin{tabular}{|l|c|c|c|}
                        \hline
                        $\Omega_{bar}$ (km s$^{-1}$ kpc$^{-1}$ )    & 31      &  41   &   51   \\
                        \hline
                        $eccentricity$                                                                                       &     $0.33\pm0.14$     &   $0.35\pm0.17$          &   $0.36\pm0.18$    \\        
                        $r_{peri}$ (kpc)                                                                                  &     $2.17\pm0.87$     &   $2.13\pm0.91$          &    $2.12\pm0.98$    \\       
                        $r_{apo}$ (kpc)                                                                                   &      $3.98\pm1.55$     &   $4.10\pm1.61$          &    $4.51\pm1.50$    \\   
                        $Z_{max}$ (kpc)                                                                                 &      $2.99\pm1.35$     &   $3.19\pm1.50$          &   $3.19\pm1.44$    \\    
                        $L_{z, min}$  ($\times$10$^{3}$ km s$^{-1}$ kpc)        &      $0.46\pm0.24$    &    $0.46\pm0.24$          &   $0.46\pm0.26$    \\                                           
                        $L_{z, max}$ ($\times$10$^{3}$ km s$^{-1}$ kpc)        &      $0.56\pm0.16$    &    $0.55\pm0.17$          &   $0.55\pm0.18$    \\                                                            
                        $E_{J}$       ($\times$10$^{5}$ km$^{2}$ s$^{-2}$)        &    $-1.96\pm0.28$     &   $-1.90\pm0.30$         &   $-1.85\pm0.32$  \\                                           
                        $E_{min}$   ($\times$10$^{5}$ km$^{2}$ s$^{-2}$)      &     $-2.17\pm0.27$    &    $-2.16\pm0.26$         &  $-2.16\pm0.25$   \\                                           
                        $E_{max}$   ($\times$10$^{5}$ km$^{2}$ s$^{-2}$)    &    $-2.07\pm0.20$     &   $-2.06\pm0.19$         &  $-2.05\pm0.17$    \\                                           
                        \hline
                \end{tabular}  \label{Table2}
        \end{center}
                \end{small}
\end{table}     

\section{Conclusions}
\label{conclusions}

We present the first high-resolution NIR spectral examination of two newly identified members of the GC Patchick~125, for which we measure precise radial velocities using data from the APOGEE-2S survey. Patchick 125 is located in a region of high interstellar reddening in the direction of the MW bulge. We employed the \texttt{BACCHUS} code to manually determine reliable abundance ratios for [Mg/Fe], [Al/Fe], [Si/Fe], and [Fe/H] from the strength of  Mg I, Al I, Si I, and Fe I lines, the intensities of which are clearly distinguished from the local continuum level (see Figure \ref{Figure2}).

Overall, Patchick~125 hosts metal-poor stars with a [Fe/H] abundance ratio of between $-1.72$ and $-1.69$, which is consistent with previous photometric estimates for this cluster. The $\alpha$ enrichment, [Mg/Fe]$\lesssim+0.20$, and in particular [Si/Fe]$\lesssim +0.30$, places Patchick~125 among typical metal-poor GCs. We identified a high enrichment in aluminum ([Al/Fe]$> +0.58$) among the Patchick~125 members, which are therefore likely associated with the so-called second-generation population. Such a value is typical among metal-poor GCs, thus reinforcing the GC nature of Patchick~125. With only two stars, it is not possible to make any meaningful conclusions about the radial velocity dispersion or metallicity spread of this cluster. 

We find that Patchick~125  is a typical metal-poor GC, which lies on a low-eccentricity and retrograde orbit and currently resides inside the corotation radius ($< 5.7$ kpc), trapped in a low-energy orbit at the edge of the Galactic bulge. 

        \begin{acknowledgements}  
                The author is grateful for the enlightening feedback from the anonymous referee. J.G.F-T acknowledges partial support from Comit\'e Mixto ESO-Chile 2021. D.M. gratefully acknowledges support from the Chilean Centro de Excelencia en Astrof\'isica y Tecnolog\'ias Afines (CATA) BASAL grant AFB-170002. ERG acknowledges support from ANID PhD scholarship No. 21210330\\

        The SDSS-IV/APOGEE-2S survey made this study possible. This work has made use of data from the European Space Agency (ESA) mission \textit{Gaia} (\url{http://www.cosmos.esa.int/gaia}), processed by the \textit{Gaia} Data Processing and Analysis Consortium (DPAC, \url{http://www.cosmos.esa.int/web/gaia/dpac/consortium}). Funding for the DPAC has been provided by national institutions, in particular the institutions participating in the \textit{Gaia} Multilateral Agreement.\\
        \end{acknowledgements}
        

\begin{thebibliography}{74}
        \expandafter\ifx\csname natexlab\endcsname\relax\def\natexlab#1{#1}\fi
        
        \bibitem[{{Ahumada} {et~al.}(2020){Ahumada}, {Prieto}, {Almeida}, {Anders},
                {Anderson}, {Andrews}, {Anguiano}, {Arcodia}, {Armengaud}, {Aubert}, {Avila},
                {Avila-Reese}, {Badenes}, {Balland}, {Barger}, {Barrera-Ballesteros}, {Basu},
                {Bautista}, {Beaton}, {Beers}, {Benavides}, {Bender}, {Bernardi}, {Bershady},
                {Beutler}, {Bidin}, {Bird}, {Bizyaev}, {Blanc}, {Blanton}, {Boquien},
                {Borissova}, {Bovy}, {Brandt}, {Brinkmann}, {Brownstein}, {Bundy}, {Bureau},
                {Burgasser}, {Burtin}, {Cano-D{\'\i}az}, {Capasso}, {Cappellari}, {Carrera},
                {Chabanier}, {Chaplin}, {Chapman}, {Cherinka}, {Chiappini}, {Doohyun Choi},
                {Chojnowski}, {Chung}, {Clerc}, {Coffey}, {Comerford}, {Comparat}, {da
                        Costa}, {Cousinou}, {Covey}, {Crane}, {Cunha}, {Ilha}, {Dai}, {Damsted},
                {Darling}, {Davidson}, {Davies}, {Dawson}, {De}, {de la Macorra}, {De Lee},
                {Queiroz}, {Deconto Machado}, {de la Torre}, {Dell'Agli}, {du Mas des
                        Bourboux}, {Diamond-Stanic}, {Dillon}, {Donor}, {Drory}, {Duckworth},
                {Dwelly}, {Ebelke}, {Eftekharzadeh}, {Davis Eigenbrot}, {Elsworth},
                {Eracleous}, {Erfanianfar}, {Escoffier}, {Fan}, {Farr},
                {Fern{\'a}ndez-Trincado}, {Feuillet}, {Finoguenov}, {Fofie},
                {Fraser-McKelvie}, {Frinchaboy}, {Fromenteau}, {Fu}, {Galbany}, {Garcia},
                {Garc{\'\i}a-Hern{\'a}ndez}, {Oehmichen}, {Ge}, {Maia}, {Geisler}, {Gelfand},
                {Goddy}, {Gonzalez-Perez}, {Grabowski}, {Green}, {Grier}, {Guo}, {Guy},
                {Harding}, {Hasselquist}, {Hawken}, {Hayes}, {Hearty}, {Hekker}, {Hogg},
                {Holtzman}, {Horta}, {Hou}, {Hsieh}, {Huber}, {Hunt}, {Chitham}, {Imig},
                {Jaber}, {Angel}, {Johnson}, {Jones}, {J{\"o}nsson}, {Jullo}, {Kim},
                {Kinemuchi}, {Kirkpatrick}, {Kite}, {Klaene}, {Kneib}, {Kollmeier}, {Kong},
                {Kounkel}, {Krishnarao}, {Lacerna}, {Lan}, {Lane}, {Law}, {Le Goff}, {Leung},
                {Lewis}, {Li}, {Lian}, {Lin}, {Long}, {Longa-Pe{\~n}a}, {Lundgren}, {Lyke},
                {Ted Mackereth}, {MacLeod}, {Majewski}, {Manchado}, {Maraston}, {Martini},
                {Masseron}, {Masters}, {Mathur}, {McDermid}, {Merloni}, {Merrifield},
                {M{\'e}sz{\'a}ros}, {Miglio}, {Minniti}, {Minsley}, {Miyaji}, {Mohammad},
                {Mosser}, {Mueller}, {Muna}, {Mu{\~n}oz-Guti{\'e}rrez}, {Myers}, {Nadathur},
                {Nair}, {Nandra}, {do Nascimento}, {Nevin}, {Newman}, {Nidever}, {Nitschelm},
                {Noterdaeme}, {O'Connell}, {Olmstead}, {Oravetz}, {Oravetz}, {Osorio},
                {Pace}, {Padilla}, {Palanque-Delabrouille}, {Palicio}, {Pan}, {Pan},
                {Parker}, {Paviot}, {Peirani}, {Ram{\'r}ez}, {Penny}, {Percival},
                {Perez-Fournon}, {P{\'e}rez-R{\`a}fols}, {Petitjean}, {Pieri},
                {Pinsonneault}, {Poovelil}, {Povick}, {Prakash}, {Price-Whelan}, {Raddick},
                {Raichoor}, {Ray}, {Rembold}, {Rezaie}, {Riffel}, {Riffel}, {Rix}, {Robin},
                {Roman-Lopes}, {Rom{\'a}n-Z{\'u}{\~n}iga}, {Rose}, {Ross}, {Rossi},
                {Rowlands}, {Rubin}, {Salvato}, {S{\'a}nchez}, {S{\'a}nchez-Menguiano},
                {S{\'a}nchez-Gallego}, {Sayres}, {Schaefer}, {Schiavon}, {Schimoia},
                {Schlafly}, {Schlegel}, {Schneider}, {Schultheis}, {Schwope}, {Seo},
                {Serenelli}, {Shafieloo}, {Shamsi}, {Shao}, {Shen}, {Shetrone}, {Shirley},
                {Aguirre}, {Simon}, {Skrutskie}, {Slosar}, {Smethurst}, {Sobeck}, {Sodi},
                {Souto}, {Stark}, {Stassun}, {Steinmetz}, {Stello}, {Stermer},
                {Storchi-Bergmann}, {Streblyanska}, {Stringfellow}, {Stutz}, {Su{\'a}rez},
                {Sun}, {Taghizadeh-Popp}, {Talbot}, {Tayar}, {Thakar}, {Theriault}, {Thomas},
                {Thomas}, {Tinker}, {Tojeiro}, {Toledo}, {Tremonti}, {Troup}, {Tuttle},
                {Unda-Sanzana}, {Valentini}, {Vargas-Gonz{\'a}lez}, {Vargas-Maga{\~n}a},
                {V{\'a}zquez-Mata}, {Vivek}, {Wake}, {Wang}, {Weaver}, {Weijmans}, {Wild},
                {Wilson}, {Wilson}, {Wolthuis}, {Wood-Vasey}, {Yan}, {Yang}, {Y{\`e}che},
                {Zamora}, {Zarrouk}, {Zasowski}, {Zhang}, {Zhao}, {Zhao}, {Zheng}, {Zheng},
                {Zhu}, \& {Zou}}]{Ahumada2020}
        {Ahumada}, R., {Prieto}, C.~A., {Almeida}, A., {et~al.} 2020, \apjs, 249, 3
        
        \bibitem[{{Alonso-Garc{\'\i}a} {et~al.}(2021){Alonso-Garc{\'\i}a}, {Smith},
                {Catelan}, {Minniti}, {Navarrete}, {Borissova}, {Carballo-Bello}, {Contreras
                        Ramos}, {Fern{\'a}ndez-Trincado}, {Ferreira Lopes}, {Gran}, {Garro},
                {Geisler}, {Guo}, {Hempel}, {Kerins}, {Lucas}, {Palma}, {Pe{\~n}a
                        Ram{\'\i}rez}, {Ram{\'\i}rez Alegr{\'\i}a}, \& {Saito}}]{Alonso-Garcia2021}
        {Alonso-Garc{\'\i}a}, J., {Smith}, L.~C., {Catelan}, M., {et~al.} 2021, \aap,
        651, A47
        
        \bibitem[{{Barb{\'a}} {et~al.}(2019){Barb{\'a}}, {Minniti}, {Geisler},
                {Alonso-Garc{\'\i}a}, {Hempel}, {Monachesi}, {Arias}, \&
                {G{\'o}mez}}]{Barba2019}
        {Barb{\'a}}, R.~H., {Minniti}, D., {Geisler}, D., {et~al.} 2019, \apjl, 870,
        L24
        
        \bibitem[{{Barbuy} {et~al.}(2021){Barbuy}, {Cantelli}, {Muniz}, {Souza},
                {Chiappini}, {Hirschi}, {Cescutti}, {Pignatari}, {Ortolani}, {Kerber},
                {Maia}, {Bica}, \& {Depagne}}]{Barbuy2021}
        {Barbuy}, B., {Cantelli}, E., {Muniz}, L., {et~al.} 2021, \aap, 654, A29
        
        \bibitem[{{Barbuy} {et~al.}(2018{\natexlab{a}}){Barbuy}, {Chiappini}, \&
                {Gerhard}}]{Barbuy2018b}
        {Barbuy}, B., {Chiappini}, C., \& {Gerhard}, O. 2018{\natexlab{a}}, \araa, 56,
        223
        
        \bibitem[{{Barbuy} {et~al.}(2018{\natexlab{b}}){Barbuy}, {Muniz}, {Ortolani},
                {Ernandes}, {Dias}, {Saviane}, {Kerber}, {Bica}, {P{\'e}rez-Villegas},
                {Rossi}, \& {Held}}]{Barbuy2018}
        {Barbuy}, B., {Muniz}, L., {Ortolani}, S., {et~al.} 2018{\natexlab{b}}, \aap,
        619, A178
        
        \bibitem[{{Bhardwaj} {et~al.}(2021){Bhardwaj}, {Rejkuba}, {Sloan}, {Marconi},
                \& {Yang}}]{Bhardwaj2021}
        {Bhardwaj}, A., {Rejkuba}, M., {Sloan}, G.~C., {Marconi}, M., \& {Yang}, S.-C.
        2021, arXiv e-prints, arXiv:2108.11388
        
        \bibitem[{{Blanton} {et~al.}(2017){Blanton}, {Bershady}, {Abolfathi},
                {Albareti}, {Allende Prieto}, {Almeida}, {Alonso-Garc{\'\i}a}, {Anders},
                {Anderson}, {Andrews}, {Aquino-Ort{\'\i}z}, {Arag{\'o}n-Salamanca},
                {Argudo-Fern{\'a}ndez}, {Armengaud}, {Aubourg}, {Avila-Reese}, {Badenes},
                {Bailey}, {Barger}, {Barrera-Ballesteros}, {Bartosz}, {Bates}, {Baumgarten},
                {Bautista}, {Beaton}, {Beers}, {Belfiore}, {Bender}, {Berlind}, {Bernardi},
                {Beutler}, {Bird}, {Bizyaev}, {Blanc}, {Blomqvist}, {Bolton}, {Boquien},
                {Borissova}, {van den Bosch}, {Bovy}, {Brandt}, {Brinkmann}, {Brownstein},
                {Bundy}, {Burgasser}, {Burtin}, {Busca}, {Cappellari}, {Delgado Carigi},
                {Carlberg}, {Carnero Rosell}, {Carrera}, {Chanover}, {Cherinka}, {Cheung},
                {G{\'o}mez Maqueo Chew}, {Chiappini}, {Choi}, {Chojnowski}, {Chuang},
                {Chung}, {Cirolini}, {Clerc}, {Cohen}, {Comparat}, {da Costa}, {Cousinou},
                {Covey}, {Crane}, {Croft}, {Cruz-Gonzalez}, {Garrido Cuadra}, {Cunha},
                {Damke}, {Darling}, {Davies}, {Dawson}, {de la Macorra}, {Dell'Agli}, {De
                        Lee}, {Delubac}, {Di Mille}, {Diamond-Stanic}, {Cano-D{\'\i}az}, {Donor},
                {Downes}, {Drory}, {du Mas des Bourboux}, {Duckworth}, {Dwelly}, {Dyer},
                {Ebelke}, {Eigenbrot}, {Eisenstein}, {Emsellem}, {Eracleous}, {Escoffier},
                {Evans}, {Fan}, {Fern{\'a}ndez-Alvar}, {Fernandez-Trincado}, {Feuillet},
                {Finoguenov}, {Fleming}, {Font-Ribera}, {Fredrickson}, {Freischlad},
                {Frinchaboy}, {Fuentes}, {Galbany}, {Garcia-Dias},
                {Garc{\'\i}a-Hern{\'a}ndez}, {Gaulme}, {Geisler}, {Gelfand},
                {Gil-Mar{\'\i}n}, {Gillespie}, {Goddard}, {Gonzalez-Perez}, {Grabowski},
                {Green}, {Grier}, {Gunn}, {Guo}, {Guy}, {Hagen}, {Hahn}, {Hall}, {Harding},
                {Hasselquist}, {Hawley}, {Hearty}, {Gonzalez Hern{\'a}ndez}, {Ho}, {Hogg},
                {Holley-Bockelmann}, {Holtzman}, {Holzer}, {Huehnerhoff}, {Hutchinson},
                {Hwang}, {Ibarra-Medel}, {da Silva Ilha}, {Ivans}, {Ivory}, {Jackson},
                {Jensen}, {Johnson}, {Jones}, {J{\"o}nsson}, {Jullo}, {Kamble}, {Kinemuchi},
                {Kirkby}, {Kitaura}, {Klaene}, {Knapp}, {Kneib}, {Kollmeier}, {Lacerna},
                {Lane}, {Lang}, {Law}, {Lazarz}, {Lee}, {Le Goff}, {Liang}, {Li}, {Li},
                {Lian}, {Lima}, {Lin}, {Lin}, {Bertran de Lis}, {Liu}, {de Icaza Lizaola},
                {Long}, {Lucatello}, {Lundgren}, {MacDonald}, {Deconto Machado}, {MacLeod},
                {Mahadevan}, {Geimba Maia}, {Maiolino}, {Majewski}, {Malanushenko},
                {Malanushenko}, {Manchado}, {Mao}, {Maraston}, {Marques-Chaves}, {Masseron},
                {Masters}, {McBride}, {McDermid}, {McGrath}, {McGreer}, {Medina Pe{\~n}a},
                {Melendez}, {Merloni}, {Merrifield}, {Meszaros}, {Meza}, {Minchev},
                {Minniti}, {Miyaji}, {More}, {Mulchaey}, {M{\"u}ller-S{\'a}nchez}, {Muna},
                {Munoz}, {Myers}, {Nair}, {Nandra}, {Correa do Nascimento}, {Negrete},
                {Ness}, {Newman}, {Nichol}, {Nidever}, {Nitschelm}, {Ntelis}, {O'Connell},
                {Oelkers}, {Oravetz}, {Oravetz}, {Pace}, {Padilla}, {Palanque-Delabrouille},
                {Alonso Palicio}, {Pan}, {Parejko}, {Parikh}, {P{\^a}ris}, {Park}, {Patten},
                {Peirani}, {Pellejero-Ibanez}, {Penny}, {Percival}, {Perez-Fournon},
                {Petitjean}, {Pieri}, {Pinsonneault}, {Pisani}, {Poleski}, {Prada},
                {Prakash}, {Queiroz}, {Raddick}, {Raichoor}, {Barboza Rembold}, {Richstein},
                {Riffel}, {Riffel}, {Rix}, {Robin}, {Rockosi}, {Rodr{\'\i}guez-Torres},
                {Roman-Lopes}, {Rom{\'a}n-Z{\'u}{\~n}iga}, {Rosado}, {Ross}, {Rossi}, {Ruan},
                {Ruggeri}, {Rykoff}, {Salazar-Albornoz}, {Salvato}, {S{\'a}nchez}, {Aguado},
                {S{\'a}nchez-Gallego}, {Santana}, {Santiago}, {Sayres}, {Schiavon}, {da Silva
                        Schimoia}, {Schlafly}, {Schlegel}, {Schneider}, {Schultheis}, {Schuster},
                {Schwope}, {Seo}, {Shao}, {Shen}, {Shetrone}, {Shull}, {Simon}, {Skinner},
                {Skrutskie}, {Slosar}, {Smith}, {Sobeck}, {Sobreira}, {Somers}, {Souto},
                {Stark}, {Stassun}, {Stauffer}, {Steinmetz}, {Storchi-Bergmann},
                {Streblyanska}, {Stringfellow}, {Su{\'a}rez}, {Sun}, {Suzuki}, {Szigeti},
                {Taghizadeh-Popp}, {Tang}, {Tao}, {Tayar}, {Tembe}, {Teske}, {Thakar},
                {Thomas}, {Thompson}, {Tinker}, {Tissera}, {Tojeiro}, {Hernandez Toledo}, {de
                        la Torre}, {Tremonti}, {Troup}, {Valenzuela}, {Martinez Valpuesta},
                {Vargas-Gonz{\'a}lez}, {Vargas-Maga{\~n}a}, {Vazquez}, {Villanova}, {Vivek},
                {Vogt}, {Wake}, {Walterbos}, {Wang}, {Weaver}, {Weijmans}, {Weinberg},
                {Westfall}, {Whelan}, {Wild}, {Wilson}, {Wood-Vasey}, {Wylezalek}, {Xiao},
                {Yan}, {Yang}, {Ybarra}, {Y{\`e}che}, {Zakamska}, {Zamora}, {Zarrouk},
                {Zasowski}, {Zhang}, {Zhao}, {Zheng}, {Zheng}, {Zhou}, {Zhou}, {Zhu},
                {Zoccali}, \& {Zou}}]{Blanton2017}
        {Blanton}, M.~R., {Bershady}, M.~A., {Abolfathi}, B., {et~al.} 2017, \aj, 154,
        28
        
        \bibitem[{{Bowen} \& {Vaughan}(1973)}]{Bowen1973}
        {Bowen}, I.~S. \& {Vaughan}, A.~H., J. 1973, \ao, 12, 1430
        
        \bibitem[{{Bressan} {et~al.}(2012){Bressan}, {Marigo}, {Girardi}, {Salasnich},
                {Dal Cero}, {Rubele}, \& {Nanni}}]{Bressan2012}
        {Bressan}, A., {Marigo}, P., {Girardi}, L., {et~al.} 2012, \mnras, 427, 127
        
        \bibitem[{{Brunthaler} {et~al.}(2011){Brunthaler}, {Reid}, {Menten}, {Zheng},
                {Bartkiewicz}, {Choi}, {Dame}, {Hachisuka}, {Immer}, {Moellenbrock},
                {Moscadelli}, {Rygl}, {Sanna}, {Sato}, {Wu}, {Xu}, \&
                {Zhang}}]{Brunthaler2011}
        {Brunthaler}, A., {Reid}, M.~J., {Menten}, K.~M., {et~al.} 2011, Astronomische
        Nachrichten, 332, 461
        
        \bibitem[{{Camargo} \& {Minniti}(2019)}]{Camargo2019}
        {Camargo}, D. \& {Minniti}, D. 2019, \mnras, 484, L90
        
        \bibitem[{{Cardelli} {et~al.}(1989){Cardelli}, {Clayton}, \& {Mathis}}]{Car89}
        {Cardelli}, J.~A., {Clayton}, G.~C., \& {Mathis}, J.~S. 1989, \apj, 345, 245
        
        \bibitem[{{Clementini} {et~al.}(2019){Clementini}, {Ripepi}, {Molinaro},
                {Garofalo}, {Muraveva}, {Rimoldini}, {Guy}, {Jevardat de Fombelle},
                {Nienartowicz}, {Marchal}, {Audard}, {Holl}, {Leccia}, {Marconi}, {Musella},
                {Mowlavi}, {Lecoeur-Taibi}, {Eyer}, {De Ridder}, {Regibo}, {Sarro},
                {Szabados}, {Evans}, \& {Riello}}]{Clementini2019}
        {Clementini}, G., {Ripepi}, V., {Molinaro}, R., {et~al.} 2019, \aap, 622, A60
        
        \bibitem[{{Contreras Ramos} {et~al.}(2018){Contreras Ramos}, {Minniti},
                {Fern{\'a}ndez-Trincado}, {Alonso-Garc{\'\i}a}, {Catelan}, {Gran}, {Hajdu},
                {Hanke}, {Hempel}, {Moreno D{\'\i}az}, {P{\'e}rez-Villegas},
                {Rojas-Arriagada}, \& {Zoccali}}]{Contreras2018}
        {Contreras Ramos}, R., {Minniti}, D., {Fern{\'a}ndez-Trincado}, J.~G., {et~al.}
        2018, \apj, 863, 78
        
        \bibitem[{{Dias} {et~al.}(2021){Dias}, {Palma}, {Minniti},
                {Fern{\'a}ndez-Trincado}, {Alonso-Garc{\'\i}a}, {Barbuy}, {Clari{\'a}},
                {Gomez}, \& {Saito}}]{Dias2021}
        {Dias}, B., {Palma}, T., {Minniti}, D., {et~al.} 2021, arXiv e-prints,
        arXiv:2110.00868

    \bibitem[Fehlberg(1968)]{F68} Fehlberg, E. 1968, NASA TR R-287      
        
        \bibitem[{{Fern{\'a}ndez-Trincado}
                {et~al.}(2021{\natexlab{a}}){Fern{\'a}ndez-Trincado}, {Beers}, {Barbuy},
                {M{\'e}sz{\'a}ros}, {Minniti}, {Smith}, {Cunha}, {Villanova}, {Geisler},
                {Majewski}, {Carigi}, {Tang}, {Moni Bidin}, \& {Vieira}}]{Trincado2021}
        {Fern{\'a}ndez-Trincado}, J.~G., {Beers}, T.~C., {Barbuy}, B., {et~al.}
        2021{\natexlab{a}}, \apjl, 918, L9
        
        \bibitem[{{Fern{\'a}ndez-Trincado}
                {et~al.}(2021{\natexlab{b}}){Fern{\'a}ndez-Trincado}, {Beers}, {Minniti},
                {Carigi}, {Placco}, {Chun}, {Lane}, {Geisler}, {Villanova}, {Souza},
                {Barbuy}, {P{\'e}rez-Villegas}, {Chiappini}, {Queiroz}, {Tang},
                {Alonso-Garc{\'\i}a}, {Piatti}, {Palma}, {Alves-Brito}, {Moni Bidin},
                {Roman-Lopes}, {Mu{\~n}oz}, {Singh}, {Kundu}, {Chaves-Velasquez},
                {Romero-Colmenares}, {Longa-Pe{\~n}a}, {Soto}, \&
                {Vieira}}]{Fernandez-Trincado2021c}
        {Fern{\'a}ndez-Trincado}, J.~G., {Beers}, T.~C., {Minniti}, D., {et~al.}
        2021{\natexlab{b}}, \aap, 647, A64
        
        \bibitem[{{Fern{\'a}ndez-Trincado}
                {et~al.}(2021{\natexlab{c}}){Fern{\'a}ndez-Trincado}, {Beers}, {Minniti},
                {Moni Bidin}, {Barbuy}, {Villanova}, {Geisler}, {Lane}, {Roman-Lopes}, \&
                {Bizyaev}}]{Fernandez-Trincad2021b}
        {Fern{\'a}ndez-Trincado}, J.~G., {Beers}, T.~C., {Minniti}, D., {et~al.}
        2021{\natexlab{c}}, \aap, 648, A70
        
        \bibitem[{{Fern{\'a}ndez-Trincado} {et~al.}(2020){Fern{\'a}ndez-Trincado},
                {Minniti}, {Beers}, {Villanova}, {Geisler}, {Souza}, {Smith}, {Placco},
                {Vieira}, {P{\'e}rez-Villegas}, {Barbuy}, {Alves-Brito}, {Bidin},
                {Alonso-Garc{\'\i}a}, {Tang}, \& {Palma}}]{Fernandez-Trincado2020}
        {Fern{\'a}ndez-Trincado}, J.~G., {Minniti}, D., {Beers}, T.~C., {et~al.} 2020,
        \aap, 643, A145
        
        \bibitem[{{Fern{\'a}ndez-Trincado}
                {et~al.}(2021{\natexlab{d}}){Fern{\'a}ndez-Trincado}, {Minniti}, {Souza},
                {Beers}, {Geisler}, {Moni Bidin}, {Villanova}, {Majewski}, {Barbuy},
                {P{\'e}rez-Villegas}, {Henao}, {Romero-Colmenares}, {Roman-Lopes}, \&
                {Lane}}]{Fernandez-Trincado2021}
        {Fern{\'a}ndez-Trincado}, J.~G., {Minniti}, D., {Souza}, S.~O., {et~al.}
        2021{\natexlab{d}}, \apjl, 908, L42
        
        \bibitem[{{Fern{\'a}ndez-Trincado}
                {et~al.}(2021{\natexlab{e}}){Fern{\'a}ndez-Trincado}, {Villanova}, {Geisler},
                {Barbuy}, {Minniti}, {Beers}, {M{\'e}sz{\'a}ros}, {Tang}, {Cohen}, {Moni
                        Bidin}, {Garro}, {Baeza}, \& {Mu{\~n}oz}}]{2021TrincadoCapos}
        {Fern{\'a}ndez-Trincado}, J.~G., {Villanova}, S., {Geisler}, D., {et~al.}
        2021{\natexlab{e}}, arXiv e-prints, arXiv:2110.10700
        
        \bibitem[{{Fern{\'a}ndez-Trincado} {et~al.}(2019){Fern{\'a}ndez-Trincado},
                {Zamora}, {Souto}, {Cohen}, {Dell'Agli}, {Garc{\'\i}a-Hern{\'a}ndez},
                {Masseron}, {Schiavon}, {M{\'e}sz{\'a}ros}, {Cunha}, {Hasselquist},
                {Shetrone}, {Schiappacasse Ulloa}, {Tang}, {Geisler}, {Schleicher},
                {Villanova}, {Mennickent}, {Minniti}, {Alonso-Garc{\'\i}a}, {Manchado},
                {Beers}, {Sobeck}, {Zasowski}, {Schultheis}, {Majewski}, {Rojas-Arriagada},
                {Almeida}, {Santana}, {Oelkers}, {Longa-Pe{\~n}a}, {Carrera}, {Burgasser},
                {Lane}, {Roman-Lopes}, {Ivans}, \& {Hearty}}]{Fernandez-Trincado2019}
        {Fern{\'a}ndez-Trincado}, J.~G., {Zamora}, O., {Souto}, D., {et~al.} 2019,
        \aap, 627, A178
        
        \bibitem[{{Gaia Collaboration} {et~al.}(2018){Gaia Collaboration}, {Brown},
                {Vallenari}, {Prusti}, {de Bruijne}, {Babusiaux}, {Bailer-Jones}, {Biermann},
                {Evans}, {Eyer}, {Jansen}, {Jordi}, {Klioner}, {Lammers}, {Lindegren},
                {Luri}, {Mignard}, {Panem}, {Pourbaix}, {Randich}, {Sartoretti}, {Siddiqui},
                {Soubiran}, {van Leeuwen}, {Walton}, {Arenou}, {Bastian}, {Cropper},
                {Drimmel}, {Katz}, {Lattanzi}, {Bakker}, {Cacciari}, {Casta{\~n}eda},
                {Chaoul}, {Cheek}, {De Angeli}, {Fabricius}, {Guerra}, {Holl}, {Masana},
                {Messineo}, {Mowlavi}, {Nienartowicz}, {Panuzzo}, {Portell}, {Riello},
                {Seabroke}, {Tanga}, {Th{\'e}venin}, {Gracia-Abril}, {Comoretto},
                {Garcia-Reinaldos}, {Teyssier}, {Altmann}, {Andrae}, {Audard},
                {Bellas-Velidis}, {Benson}, {Berthier}, {Blomme}, {Burgess}, {Busso},
                {Carry}, {Cellino}, {Clementini}, {Clotet}, {Creevey}, {Davidson}, {De
                        Ridder}, {Delchambre}, {Dell'Oro}, {Ducourant},
                {Fern{\'a}ndez-Hern{\'a}ndez}, {Fouesneau}, {Fr{\'e}mat}, {Galluccio},
                {Garc{\'\i}a-Torres}, {Gonz{\'a}lez-N{\'u}{\~n}ez}, {Gonz{\'a}lez-Vidal},
                {Gosset}, {Guy}, {Halbwachs}, {Hambly}, {Harrison}, {Hern{\'a}ndez},
                {Hestroffer}, {Hodgkin}, {Hutton}, {Jasniewicz}, {Jean-Antoine-Piccolo},
                {Jordan}, {Korn}, {Krone-Martins}, {Lanzafame}, {Lebzelter}, {L{\"o}ffler},
                {Manteiga}, {Marrese}, {Mart{\'\i}n-Fleitas}, {Moitinho}, {Mora}, {Muinonen},
                {Osinde}, {Pancino}, {Pauwels}, {Petit}, {Recio-Blanco}, {Richards},
                {Rimoldini}, {Robin}, {Sarro}, {Siopis}, {Smith}, {Sozzetti}, {S{\"u}veges},
                {Torra}, {van Reeven}, {Abbas}, {Abreu Aramburu}, {Accart}, {Aerts},
                {Altavilla}, {{\'A}lvarez}, {Alvarez}, {Alves}, {Anderson}, {Andrei},
                {Anglada Varela}, {Antiche}, {Antoja}, {Arcay}, {Astraatmadja}, {Bach},
                {Baker}, {Balaguer-N{\'u}{\~n}ez}, {Balm}, {Barache}, {Barata}, {Barbato},
                {Barblan}, {Barklem}, {Barrado}, {Barros}, {Barstow}, {Bartholom{\'e}
                        Mu{\~n}oz}, {Bassilana}, {Becciani}, {Bellazzini}, {Berihuete}, {Bertone},
                {Bianchi}, {Bienaym{\'e}}, {Blanco-Cuaresma}, {Boch}, {Boeche}, {Bombrun},
                {Borrachero}, {Bossini}, {Bouquillon}, {Bourda}, {Bragaglia}, {Bramante},
                {Breddels}, {Bressan}, {Brouillet}, {Br{\"u}semeister}, {Brugaletta},
                {Bucciarelli}, {Burlacu}, {Busonero}, {Butkevich}, {Buzzi}, {Caffau},
                {Cancelliere}, {Cannizzaro}, {Cantat-Gaudin}, {Carballo}, {Carlucci},
                {Carrasco}, {Casamiquela}, {Castellani}, {Castro-Ginard}, {Charlot},
                {Chemin}, {Chiavassa}, {Cocozza}, {Costigan}, {Cowell}, {Crifo}, {Crosta},
                {Crowley}, {Cuypers}, {Dafonte}, {Damerdji}, {Dapergolas}, {David}, {David},
                {de Laverny}, {De Luise}, {De March}, {de Martino}, {de Souza}, {de Torres},
                {Debosscher}, {del Pozo}, {Delbo}, {Delgado}, {Delgado}, {Di Matteo},
                {Diakite}, {Diener}, {Distefano}, {Dolding}, {Drazinos}, {Dur{\'a}n},
                {Edvardsson}, {Enke}, {Eriksson}, {Esquej}, {Eynard Bontemps}, {Fabre},
                {Fabrizio}, {Faigler}, {Falc{\~a}o}, {Farr{\`a}s Casas}, {Federici},
                {Fedorets}, {Fernique}, {Figueras}, {Filippi}, {Findeisen}, {Fonti},
                {Fraile}, {Fraser}, {Fr{\'e}zouls}, {Gai}, {Galleti}, {Garabato},
                {Garc{\'\i}a-Sedano}, {Garofalo}, {Garralda}, {Gavel}, {Gavras}, {Gerssen},
                {Geyer}, {Giacobbe}, {Gilmore}, {Girona}, {Giuffrida}, {Glass}, {Gomes},
                {Granvik}, {Gueguen}, {Guerrier}, {Guiraud}, {Guti{\'e}rrez-S{\'a}nchez},
                {Haigron}, {Hatzidimitriou}, {Hauser}, {Haywood}, {Heiter}, {Helmi}, {Heu},
                {Hilger}, {Hobbs}, {Hofmann}, {Holland}, {Huckle}, {Hypki}, {Icardi},
                {Jan{\ss}en}, {Jevardat de Fombelle}, {Jonker}, {Juh{\'a}sz}, {Julbe},
                {Karampelas}, {Kewley}, {Klar}, {Kochoska}, {Kohley}, {Kolenberg},
                {Kontizas}, {Kontizas}, {Koposov}, {Kordopatis}, {Kostrzewa-Rutkowska},
                {Koubsky}, {Lambert}, {Lanza}, {Lasne}, {Lavigne}, {Le Fustec}, {Le
                        Poncin-Lafitte}, {Lebreton}, {Leccia}, {Leclerc}, {Lecoeur-Taibi},
                {Lenhardt}, {Leroux}, {Liao}, {Licata}, {Lindstr{\o}m}, {Lister}, {Livanou},
                {Lobel}, {L{\'o}pez}, {Managau}, {Mann}, {Mantelet}, {Marchal}, {Marchant},
                {Marconi}, {Marinoni}, {Marschalk{\'o}}, {Marshall}, {Martino}, {Marton},
                {Mary}, {Massari}, {Matijevi{\v{c}}}, {Mazeh}, {McMillan}, {Messina},
                {Michalik}, {Millar}, {Molina}, {Molinaro}, {Moln{\'a}r}, {Montegriffo},
                {Mor}, {Morbidelli}, {Morel}, {Morris}, {Mulone}, {Muraveva}, {Musella},
                {Nelemans}, {Nicastro}, {Noval}, {O'Mullane}, {Ord{\'e}novic},
                {Ord{\'o}{\~n}ez-Blanco}, {Osborne}, {Pagani}, {Pagano}, {Pailler},
                {Palacin}, {Palaversa}, {Panahi}, {Pawlak}, {Piersimoni}, {Pineau}, {Plachy},
                {Plum}, {Poggio}, {Poujoulet}, {Pr{\v{s}}a}, {Pulone}, {Racero}, {Ragaini},
                {Rambaux}, {Ramos-Lerate}, {Regibo}, {Reyl{\'e}}, {Riclet}, {Ripepi}, {Riva},
                {Rivard}, {Rixon}, {Roegiers}, {Roelens}, {Romero-G{\'o}mez}, {Rowell},
                {Royer}, {Ruiz-Dern}, {Sadowski}, {Sagrist{\`a} Sell{\'e}s}, {Sahlmann},
                {Salgado}, {Salguero}, {Sanna}, {Santana-Ros}, {Sarasso}, {Savietto},
                {Schultheis}, {Sciacca}, {Segol}, {Segovia}, {S{\'e}gransan}, {Shih},
                {Siltala}, {Silva}, {Smart}, {Smith}, {Solano}, {Solitro}, {Sordo}, {Soria
                        Nieto}, {Souchay}, {Spagna}, {Spoto}, {Stampa}, {Steele},
                {Steidelm{\"u}ller}, {Stephenson}, {Stoev}, {Suess}, {Surdej}, {Szabados},
                {Szegedi-Elek}, {Tapiador}, {Taris}, {Tauran}, {Taylor}, {Teixeira},
                {Terrett}, {Teyssandier}, {Thuillot}, {Titarenko}, {Torra Clotet}, {Turon},
                {Ulla}, {Utrilla}, {Uzzi}, {Vaillant}, {Valentini}, {Valette}, {van Elteren},
                {Van Hemelryck}, {van Leeuwen}, {Vaschetto}, {Vecchiato}, {Veljanoski},
                {Viala}, {Vicente}, {Vogt}, {von Essen}, {Voss}, {Votruba}, {Voutsinas},
                {Walmsley}, {Weiler}, {Wertz}, {Wevers}, {Wyrzykowski}, {Yoldas},
                {{\v{Z}}erjal}, {Ziaeepour}, {Zorec}, {Zschocke}, {Zucker}, {Zurbach}, \&
                {Zwitter}}]{Gaia2018}
        {Gaia Collaboration}, {Brown}, A.~G.~A., {Vallenari}, A., {et~al.} 2018, \aap,
        616, A1
        
        \bibitem[{{Gaia Collaboration} {et~al.}(2021){Gaia Collaboration}, {Brown},
                {Vallenari}, {Prusti}, {de Bruijne}, {Babusiaux}, {Biermann}, {Creevey},
                {Evans}, {Eyer}, {Hutton}, {Jansen}, {Jordi}, {Klioner}, {Lammers},
                {Lindegren}, {Luri}, {Mignard}, {Panem}, {Pourbaix}, {Randich}, {Sartoretti},
                {Soubiran}, {Walton}, {Arenou}, {Bailer-Jones}, {Bastian}, {Cropper},
                {Drimmel}, {Katz}, {Lattanzi}, {van Leeuwen}, {Bakker}, {Cacciari},
                {Casta{\~n}eda}, {De Angeli}, {Ducourant}, {Fabricius}, {Fouesneau},
                {Fr{\'e}mat}, {Guerra}, {Guerrier}, {Guiraud}, {Jean-Antoine Piccolo},
                {Masana}, {Messineo}, {Mowlavi}, {Nicolas}, {Nienartowicz}, {Pailler},
                {Panuzzo}, {Riclet}, {Roux}, {Seabroke}, {Sordo}, {Tanga}, {Th{\'e}venin},
                {Gracia-Abril}, {Portell}, {Teyssier}, {Altmann}, {Andrae}, {Bellas-Velidis},
                {Benson}, {Berthier}, {Blomme}, {Brugaletta}, {Burgess}, {Busso}, {Carry},
                {Cellino}, {Cheek}, {Clementini}, {Damerdji}, {Davidson}, {Delchambre},
                {Dell'Oro}, {Fern{\'a}ndez-Hern{\'a}ndez}, {Galluccio}, {Garc{\'\i}a-Lario},
                {Garcia-Reinaldos}, {Gonz{\'a}lez-N{\'u}{\~n}ez}, {Gosset}, {Haigron},
                {Halbwachs}, {Hambly}, {Harrison}, {Hatzidimitriou}, {Heiter},
                {Hern{\'a}ndez}, {Hestroffer}, {Hodgkin}, {Holl}, {Jan{\ss}en}, {Jevardat de
                        Fombelle}, {Jordan}, {Krone-Martins}, {Lanzafame}, {L{\"o}ffler}, {Lorca},
                {Manteiga}, {Marchal}, {Marrese}, {Moitinho}, {Mora}, {Muinonen}, {Osborne},
                {Pancino}, {Pauwels}, {Petit}, {Recio-Blanco}, {Richards}, {Riello},
                {Rimoldini}, {Robin}, {Roegiers}, {Rybizki}, {Sarro}, {Siopis}, {Smith},
                {Sozzetti}, {Ulla}, {Utrilla}, {van Leeuwen}, {van Reeven}, {Abbas}, {Abreu
                        Aramburu}, {Accart}, {Aerts}, {Aguado}, {Ajaj}, {Altavilla}, {{\'A}lvarez},
                {{\'A}lvarez Cid-Fuentes}, {Alves}, {Anderson}, {Anglada Varela}, {Antoja},
                {Audard}, {Baines}, {Baker}, {Balaguer-N{\'u}{\~n}ez}, {Balbinot}, {Balog},
                {Barache}, {Barbato}, {Barros}, {Barstow}, {Bartolom{\'e}}, {Bassilana},
                {Bauchet}, {Baudesson-Stella}, {Becciani}, {Bellazzini}, {Bernet}, {Bertone},
                {Bianchi}, {Blanco-Cuaresma}, {Boch}, {Bombrun}, {Bossini}, {Bouquillon},
                {Bragaglia}, {Bramante}, {Breedt}, {Bressan}, {Brouillet}, {Bucciarelli},
                {Burlacu}, {Busonero}, {Butkevich}, {Buzzi}, {Caffau}, {Cancelliere},
                {C{\'a}novas}, {Cantat-Gaudin}, {Carballo}, {Carlucci}, {Carnerero},
                {Carrasco}, {Casamiquela}, {Castellani}, {Castro-Ginard}, {Castro Sampol},
                {Chaoul}, {Charlot}, {Chemin}, {Chiavassa}, {Cioni}, {Comoretto}, {Cooper},
                {Cornez}, {Cowell}, {Crifo}, {Crosta}, {Crowley}, {Dafonte}, {Dapergolas},
                {David}, {David}, {de Laverny}, {De Luise}, {De March}, {De Ridder}, {de
                        Souza}, {de Teodoro}, {de Torres}, {del Peloso}, {del Pozo}, {Delbo},
                {Delgado}, {Delgado}, {Delisle}, {Di Matteo}, {Diakite}, {Diener},
                {Distefano}, {Dolding}, {Eappachen}, {Edvardsson}, {Enke}, {Esquej}, {Fabre},
                {Fabrizio}, {Faigler}, {Fedorets}, {Fernique}, {Fienga}, {Figueras},
                {Fouron}, {Fragkoudi}, {Fraile}, {Franke}, {Gai}, {Garabato},
                {Garcia-Gutierrez}, {Garc{\'\i}a-Torres}, {Garofalo}, {Gavras}, {Gerlach},
                {Geyer}, {Giacobbe}, {Gilmore}, {Girona}, {Giuffrida}, {Gomel}, {Gomez},
                {Gonzalez-Santamaria}, {Gonz{\'a}lez-Vidal}, {Granvik},
                {Guti{\'e}rrez-S{\'a}nchez}, {Guy}, {Hauser}, {Haywood}, {Helmi}, {Hidalgo},
                {Hilger}, {H{\l}adczuk}, {Hobbs}, {Holland}, {Huckle}, {Jasniewicz},
                {Jonker}, {Juaristi Campillo}, {Julbe}, {Karbevska}, {Kervella}, {Khanna},
                {Kochoska}, {Kontizas}, {Kordopatis}, {Korn}, {Kostrzewa-Rutkowska},
                {Kruszy{\'n}ska}, {Lambert}, {Lanza}, {Lasne}, {Le Campion}, {Le Fustec},
                {Lebreton}, {Lebzelter}, {Leccia}, {Leclerc}, {Lecoeur-Taibi}, {Liao},
                {Licata}, {Lindstr{\o}m}, {Lister}, {Livanou}, {Lobel}, {Madrero Pardo},
                {Managau}, {Mann}, {Marchant}, {Marconi}, {Marcos Santos}, {Marinoni},
                {Marocco}, {Marshall}, {Martin Polo}, {Mart{\'\i}n-Fleitas}, {Masip},
                {Massari}, {Mastrobuono-Battisti}, {Mazeh}, {McMillan}, {Messina},
                {Michalik}, {Millar}, {Mints}, {Molina}, {Molinaro}, {Moln{\'a}r},
                {Montegriffo}, {Mor}, {Morbidelli}, {Morel}, {Morris}, {Mulone}, {Munoz},
                {Muraveva}, {Murphy}, {Musella}, {Noval}, {Ord{\'e}novic}, {Orr{\`u}},
                {Osinde}, {Pagani}, {Pagano}, {Palaversa}, {Palicio}, {Panahi}, {Pawlak},
                {Pe{\~n}alosa Esteller}, {Penttil{\"a}}, {Piersimoni}, {Pineau}, {Plachy},
                {Plum}, {Poggio}, {Poretti}, {Poujoulet}, {Pr{\v{s}}a}, {Pulone}, {Racero},
                {Ragaini}, {Rainer}, {Raiteri}, {Rambaux}, {Ramos}, {Ramos-Lerate}, {Re
                        Fiorentin}, {Regibo}, {Reyl{\'e}}, {Ripepi}, {Riva}, {Rixon}, {Robichon},
                {Robin}, {Roelens}, {Rohrbasser}, {Romero-G{\'o}mez}, {Rowell}, {Royer},
                {Rybicki}, {Sadowski}, {Sagrist{\`a} Sell{\'e}s}, {Sahlmann}, {Salgado},
                {Salguero}, {Samaras}, {Sanchez Gimenez}, {Sanna}, {Santove{\~n}a},
                {Sarasso}, {Schultheis}, {Sciacca}, {Segol}, {Segovia}, {S{\'e}gransan},
                {Semeux}, {Shahaf}, {Siddiqui}, {Siebert}, {Siltala}, {Slezak}, {Smart},
                {Solano}, {Solitro}, {Souami}, {Souchay}, {Spagna}, {Spoto}, {Steele},
                {Steidelm{\"u}ller}, {Stephenson}, {S{\"u}veges}, {Szabados}, {Szegedi-Elek},
                {Taris}, {Tauran}, {Taylor}, {Teixeira}, {Thuillot}, {Tonello}, {Torra},
                {Torra}, {Turon}, {Unger}, {Vaillant}, {van Dillen}, {Vanel}, {Vecchiato},
                {Viala}, {Vicente}, {Voutsinas}, {Weiler}, {Wevers}, {Wyrzykowski}, {Yoldas},
                {Yvard}, {Zhao}, {Zorec}, {Zucker}, {Zurbach}, \& {Zwitter}}]{Gaia2021}
        {Gaia Collaboration}, {Brown}, A.~G.~A., {Vallenari}, A., {et~al.} 2021, \aap,
        650, C3
        
        \bibitem[{{Gaia Collaboration} {et~al.}(2016){Gaia Collaboration}, {Brown},
                {Vallenari}, {Prusti}, {de Bruijne}, {Mignard}, {Drimmel}, {Babusiaux},
                {Bailer-Jones}, {Bastian}, {Biermann}, {Evans}, {Eyer}, {Jansen}, {Jordi},
                {Katz}, {Klioner}, {Lammers}, {Lindegren}, {Luri}, {O'Mullane}, {Panem},
                {Pourbaix}, {Randich}, {Sartoretti}, {Siddiqui}, {Soubiran}, {Valette}, {van
                        Leeuwen}, {Walton}, {Aerts}, {Arenou}, {Cropper}, {H{\o}g}, {Lattanzi},
                {Grebel}, {Holland}, {Huc}, {Passot}, {Perryman}, {Bramante}, {Cacciari},
                {Casta{\~n}eda}, {Chaoul}, {Cheek}, {De Angeli}, {Fabricius}, {Guerra},
                {Hern{\'a}ndez}, {Jean-Antoine-Piccolo}, {Masana}, {Messineo}, {Mowlavi},
                {Nienartowicz}, {Ord{\'o}{\~n}ez-Blanco}, {Panuzzo}, {Portell}, {Richards},
                {Riello}, {Seabroke}, {Tanga}, {Th{\'e}venin}, {Torra}, {Els},
                {Gracia-Abril}, {Comoretto}, {Garcia-Reinaldos}, {Lock}, {Mercier},
                {Altmann}, {Andrae}, {Astraatmadja}, {Bellas-Velidis}, {Benson}, {Berthier},
                {Blomme}, {Busso}, {Carry}, {Cellino}, {Clementini}, {Cowell}, {Creevey},
                {Cuypers}, {Davidson}, {De Ridder}, {de Torres}, {Delchambre}, {Dell'Oro},
                {Ducourant}, {Fr{\'e}mat}, {Garc{\'\i}a-Torres}, {Gosset}, {Halbwachs},
                {Hambly}, {Harrison}, {Hauser}, {Hestroffer}, {Hodgkin}, {Huckle}, {Hutton},
                {Jasniewicz}, {Jordan}, {Kontizas}, {Korn}, {Lanzafame}, {Manteiga},
                {Moitinho}, {Muinonen}, {Osinde}, {Pancino}, {Pauwels}, {Petit},
                {Recio-Blanco}, {Robin}, {Sarro}, {Siopis}, {Smith}, {Smith}, {Sozzetti},
                {Thuillot}, {van Reeven}, {Viala}, {Abbas}, {Abreu Aramburu}, {Accart},
                {Aguado}, {Allan}, {Allasia}, {Altavilla}, {{\'A}lvarez}, {Alves},
                {Anderson}, {Andrei}, {Anglada Varela}, {Antiche}, {Antoja}, {Ant{\'o}n},
                {Arcay}, {Bach}, {Baker}, {Balaguer-N{\'u}{\~n}ez}, {Barache}, {Barata},
                {Barbier}, {Barblan}, {Barrado y Navascu{\'e}s}, {Barros}, {Barstow},
                {Becciani}, {Bellazzini}, {Bello Garc{\'\i}a}, {Belokurov}, {Bendjoya},
                {Berihuete}, {Bianchi}, {Bienaym{\'e}}, {Billebaud}, {Blagorodnova},
                {Blanco-Cuaresma}, {Boch}, {Bombrun}, {Borrachero}, {Bouquillon}, {Bourda},
                {Bouy}, {Bragaglia}, {Breddels}, {Brouillet}, {Br{\"u}semeister},
                {Bucciarelli}, {Burgess}, {Burgon}, {Burlacu}, {Busonero}, {Buzzi}, {Caffau},
                {Cambras}, {Campbell}, {Cancelliere}, {Cantat-Gaudin}, {Carlucci},
                {Carrasco}, {Castellani}, {Charlot}, {Charnas}, {Chiavassa}, {Clotet},
                {Cocozza}, {Collins}, {Costigan}, {Crifo}, {Cross}, {Crosta}, {Crowley},
                {Dafonte}, {Damerdji}, {Dapergolas}, {David}, {David}, {De Cat}, {de Felice},
                {de Laverny}, {De Luise}, {De March}, {de Martino}, {de Souza}, {Debosscher},
                {del Pozo}, {Delbo}, {Delgado}, {Delgado}, {Di Matteo}, {Diakite},
                {Distefano}, {Dolding}, {Dos Anjos}, {Drazinos}, {Duran}, {Dzigan},
                {Edvardsson}, {Enke}, {Evans}, {Eynard Bontemps}, {Fabre}, {Fabrizio},
                {Faigler}, {Falc{\~a}o}, {Farr{\`a}s Casas}, {Federici}, {Fedorets},
                {Fern{\'a}ndez-Hern{\'a}ndez}, {Fernique}, {Fienga}, {Figueras}, {Filippi},
                {Findeisen}, {Fonti}, {Fouesneau}, {Fraile}, {Fraser}, {Fuchs}, {Gai},
                {Galleti}, {Galluccio}, {Garabato}, {Garc{\'\i}a-Sedano}, {Garofalo},
                {Garralda}, {Gavras}, {Gerssen}, {Geyer}, {Gilmore}, {Girona}, {Giuffrida},
                {Gomes}, {Gonz{\'a}lez-Marcos}, {Gonz{\'a}lez-N{\'u}{\~n}ez},
                {Gonz{\'a}lez-Vidal}, {Granvik}, {Guerrier}, {Guillout}, {Guiraud},
                {G{\'u}rpide}, {Guti{\'e}rrez-S{\'a}nchez}, {Guy}, {Haigron},
                {Hatzidimitriou}, {Haywood}, {Heiter}, {Helmi}, {Hobbs}, {Hofmann}, {Holl},
                {Holland}, {Hunt}, {Hypki}, {Icardi}, {Irwin}, {Jevardat de Fombelle},
                {Jofr{\'e}}, {Jonker}, {Jorissen}, {Julbe}, {Karampelas}, {Kochoska},
                {Kohley}, {Kolenberg}, {Kontizas}, {Koposov}, {Kordopatis}, {Koubsky},
                {Krone-Martins}, {Kudryashova}, {Kull}, {Bachchan}, {Lacoste-Seris}, {Lanza},
                {Lavigne}, {Le Poncin-Lafitte}, {Lebreton}, {Lebzelter}, {Leccia}, {Leclerc},
                {Lecoeur-Taibi}, {Lemaitre}, {Lenhardt}, {Leroux}, {Liao}, {Licata},
                {Lindstr{\o}m}, {Lister}, {Livanou}, {Lobel}, {L{\"o}ffler}, {L{\'o}pez},
                {Lorenz}, {MacDonald}, {Magalh{\~a}es Fernandes}, {Managau}, {Mann},
                {Mantelet}, {Marchal}, {Marchant}, {Marconi}, {Marinoni}, {Marrese},
                {Marschalk{\'o}}, {Marshall}, {Mart{\'\i}n-Fleitas}, {Martino}, {Mary},
                {Matijevi{\v{c}}}, {Mazeh}, {McMillan}, {Messina}, {Michalik}, {Millar},
                {Miranda}, {Molina}, {Molinaro}, {Molinaro}, {Moln{\'a}r}, {Moniez},
                {Montegriffo}, {Mor}, {Mora}, {Morbidelli}, {Morel}, {Morgenthaler},
                {Morris}, {Mulone}, {Muraveva}, {Musella}, {Narbonne}, {Nelemans},
                {Nicastro}, {Noval}, {Ord{\'e}novic}, {Ordieres-Mer{\'e}}, {Osborne},
                {Pagani}, {Pagano}, {Pailler}, {Palacin}, {Palaversa}, {Parsons}, {Pecoraro},
                {Pedrosa}, {Pentik{\"a}inen}, {Pichon}, {Piersimoni}, {Pineau}, {Plachy},
                {Plum}, {Poujoulet}, {Pr{\v{s}}a}, {Pulone}, {Ragaini}, {Rago}, {Rambaux},
                {Ramos-Lerate}, {Ranalli}, {Rauw}, {Read}, {Regibo}, {Reyl{\'e}}, {Ribeiro},
                {Rimoldini}, {Ripepi}, {Riva}, {Rixon}, {Roelens}, {Romero-G{\'o}mez},
                {Rowell}, {Royer}, {Ruiz-Dern}, {Sadowski}, {Sagrist{\`a} Sell{\'e}s},
                {Sahlmann}, {Salgado}, {Salguero}, {Sarasso}, {Savietto}, {Schultheis},
                {Sciacca}, {Segol}, {Segovia}, {Segransan}, {Shih}, {Smareglia}, {Smart},
                {Solano}, {Solitro}, {Sordo}, {Soria Nieto}, {Souchay}, {Spagna}, {Spoto},
                {Stampa}, {Steele}, {Steidelm{\"u}ller}, {Stephenson}, {Stoev}, {Suess},
                {S{\"u}veges}, {Surdej}, {Szabados}, {Szegedi-Elek}, {Tapiador}, {Taris},
                {Tauran}, {Taylor}, {Teixeira}, {Terrett}, {Tingley}, {Trager}, {Turon},
                {Ulla}, {Utrilla}, {Valentini}, {van Elteren}, {Van Hemelryck}, {van
                        Leeuwen}, {Varadi}, {Vecchiato}, {Veljanoski}, {Via}, {Vicente}, {Vogt},
                {Voss}, {Votruba}, {Voutsinas}, {Walmsley}, {Weiler}, {Weingrill}, {Wevers},
                {Wyrzykowski}, {Yoldas}, {{\v{Z}}erjal}, {Zucker}, {Zurbach}, {Zwitter},
                {Alecu}, {Allen}, {Allende Prieto}, {Amorim}, {Anglada-Escud{\'e}},
                {Arsenijevic}, {Azaz}, {Balm}, {Beck}, {Bernstein}, {Bigot}, {Bijaoui},
                {Blasco}, {Bonfigli}, {Bono}, {Boudreault}, {Bressan}, {Brown}, {Brunet},
                {Bunclark}, {Buonanno}, {Butkevich}, {Carret}, {Carrion}, {Chemin},
                {Ch{\'e}reau}, {Corcione}, {Darmigny}, {de Boer}, {de Teodoro}, {de Zeeuw},
                {Delle Luche}, {Domingues}, {Dubath}, {Fodor}, {Fr{\'e}zouls}, {Fries},
                {Fustes}, {Fyfe}, {Gallardo}, {Gallegos}, {Gardiol}, {Gebran}, {Gomboc},
                {G{\'o}mez}, {Grux}, {Gueguen}, {Heyrovsky}, {Hoar}, {Iannicola}, {Isasi
                        Parache}, {Janotto}, {Joliet}, {Jonckheere}, {Keil}, {Kim}, {Klagyivik},
                {Klar}, {Knude}, {Kochukhov}, {Kolka}, {Kos}, {Kutka}, {Lainey}, {LeBouquin},
                {Liu}, {Loreggia}, {Makarov}, {Marseille}, {Martayan}, {Martinez-Rubi},
                {Massart}, {Meynadier}, {Mignot}, {Munari}, {Nguyen}, {Nordlander}, {Ocvirk},
                {O'Flaherty}, {Olias Sanz}, {Ortiz}, {Osorio}, {Oszkiewicz}, {Ouzounis},
                {Palmer}, {Park}, {Pasquato}, {Peltzer}, {Peralta}, {P{\'e}turaud},
                {Pieniluoma}, {Pigozzi}, {Poels}, {Prat}, {Prod'homme}, {Raison}, {Rebordao},
                {Risquez}, {Rocca-Volmerange}, {Rosen}, {Ruiz-Fuertes}, {Russo}, {Sembay},
                {Serraller Vizcaino}, {Short}, {Siebert}, {Silva}, {Sinachopoulos}, {Slezak},
                {Soffel}, {Sosnowska}, {Strai{\v{z}}ys}, {ter Linden}, {Terrell}, {Theil},
                {Tiede}, {Troisi}, {Tsalmantza}, {Tur}, {Vaccari}, {Vachier}, {Valles}, {Van
                        Hamme}, {Veltz}, {Virtanen}, {Wallut}, {Wichmann}, {Wilkinson}, {Ziaeepour},
                \& {Zschocke}}]{Gaia2016}
        {Gaia Collaboration}, {Brown}, A.~G.~A., {Vallenari}, A., {et~al.} 2016, \aap,
        595, A2
        
        \bibitem[{{Garc{\'\i}a P{\'e}rez} {et~al.}(2016){Garc{\'\i}a P{\'e}rez},
                {Allende Prieto}, {Holtzman}, {Shetrone}, {M{\'e}sz{\'a}ros}, {Bizyaev},
                {Carrera}, {Cunha}, {Garc{\'\i}a-Hern{\'a}ndez}, {Johnson}, {Majewski},
                {Nidever}, {Schiavon}, {Shane}, {Smith}, {Sobeck}, {Troup}, {Zamora},
                {Weinberg}, {Bovy}, {Eisenstein}, {Feuillet}, {Frinchaboy}, {Hayden},
                {Hearty}, {Nguyen}, {O'Connell}, {Pinsonneault}, {Wilson}, \&
                {Zasowski}}]{Garcia2016}
        {Garc{\'\i}a P{\'e}rez}, A.~E., {Allende Prieto}, C., {Holtzman}, J.~A.,
        {et~al.} 2016, \aj, 151, 144
        
        \bibitem[{{Garro} {et~al.}(2021{\natexlab{a}}){Garro}, {Minniti}, {G{\'o}mez},
                \& {Alonso-Garc{\'\i}a}}]{Garro2021b}
        {Garro}, E.~R., {Minniti}, D., {G{\'o}mez}, M., \& {Alonso-Garc{\'\i}a}, J.
        2021{\natexlab{a}}, \aap, 654, A23
        
        \bibitem[{{Garro} {et~al.}(2020){Garro}, {Minniti}, {G{\'o}mez},
                {Alonso-Garc{\'\i}a}, {Barb{\'a}}, {Barbuy}, {Clari{\'a}}, {Chen{\'e}},
                {Dias}, {Hempel}, {Ivanov}, {Lucas}, {Majaess}, {Mauro}, {Moni Bidin},
                {Palma}, {Pullen}, {Saito}, {Smith}, {Surot}, {Ram{\'\i}rez Alegr{\'\i}a},
                {Rejkuba}, {Ripepi}, \& {Fern{\'a}ndez Trincado}}]{Garro2020}
        {Garro}, E.~R., {Minniti}, D., {G{\'o}mez}, M., {et~al.} 2020, \aap, 642, L19
        
        \bibitem[{{Garro} {et~al.}(2021{\natexlab{b}}){Garro}, {Minniti}, {G{\'o}mez},
                {Alonso-Garc{\'\i}a}, {Palma}, {Smith}, \& {Ripepi}}]{Garro2021a}
        {Garro}, E.~R., {Minniti}, D., {G{\'o}mez}, M., {et~al.} 2021{\natexlab{b}},
        \aap, 649, A86
        
        \bibitem[{{Garro} {et~al.}(2021{\natexlab{c}}){Garro}, {Minniti}, {G{\'o}mez},
                {Alonso-Garc{\'\i}a}, {Ripepi}, {Fern\'andez-Trincado}, \& {Vivanco
                        C\'adiz}}]{Garro2021c}
        {Garro}, E.~R., {Minniti}, D., {G{\'o}mez}, M., {et~al.} 2021{\natexlab{c}},
        \aap, submitted
        
        \bibitem[{{Geisler} {et~al.}(2021){Geisler}, {Villanova}, {O'Connell}, {Cohen},
                {Moni Bidin}, {Fern{\'a}ndez-Trincado}, {Mu{\~n}oz}, {Minniti}, {Zoccali},
                {Rojas-Arriagada}, {Contreras Ramos}, {Catelan}, {Mauro}, {Cort{\'e}s},
                {Ferreira Lopes}, {Arentsen}, {Starkenburg}, {Martin}, {Tang}, {Parisi},
                {Alonso-Garc{\'\i}a}, {Gran}, {Cunha}, {Smith}, {Majewski}, {J{\"o}nsson},
                {Garc{\'\i}a-Hern{\'a}ndez}, {Horta}, {M{\'e}sz{\'a}ros}, {Monaco},
                {Monachesi}, {Mu{\~n}oz}, {Brownstein}, {Beers}, {Lane}, {Barbuy}, {Sobeck},
                {Henao}, {Gonz{\'a}lez-D{\'\i}az}, {Miranda}, {Reinarz}, \&
                {Santander}}]{Geisler2021}
        {Geisler}, D., {Villanova}, S., {O'Connell}, J.~E., {et~al.} 2021, \aap, 652,
        A157
        
        \bibitem[{{Gran} {et~al.}(2019){Gran}, {Zoccali}, {Contreras Ramos}, {Valenti},
                {Rojas-Arriagada}, {Carballo-Bello}, {Alonso-Garcia}, {Minniti}, {Rejkuba},
                \& {Surot}}]{Gran2019}
        {Gran}, F., {Zoccali}, M., {Contreras Ramos}, R., {et~al.} 2019, \aap, 628, A45
        
        \bibitem[{{Gran} {et~al.}(2021{\natexlab{a}}){Gran}, {Zoccali},
                {Rojas-Arriagada}, {Saviane}, {Contreras Ramos}, {Beaton}, {Bizyaev},
                {Cohen}, {Fern{\'a}ndez-Trincado}, {Garc{\'\i}a-Hern{\'a}ndez}, {Geisler},
                {Lane}, {Minniti}, {Moni Bidin}, {Nitschelm}, {Olivares Carvajal}, {Pan},
                {Rojas}, \& {Villanova}}]{Gran2021a}
        {Gran}, F., {Zoccali}, M., {Rojas-Arriagada}, A., {et~al.} 2021{\natexlab{a}},
        \mnras, 504, 3494
        
        \bibitem[{{Gran} {et~al.}(2021{\natexlab{b}}){Gran}, {Zoccali}, {Saviane},
                {Valenti}, {Rojas-Arriagada}, {Ramos}, {Hartke}, {Carballo-Bello},
                {Navarrete}, {Rejkuba}, \& {Carvajal}}]{Gran2021b}
        {Gran}, F., {Zoccali}, M., {Saviane}, I., {et~al.} 2021{\natexlab{b}}, \mnras
        [\eprint[arXiv]{2108.11922}]
        
        \bibitem[{{Gustafsson} {et~al.}(2008){Gustafsson}, {Edvardsson}, {Eriksson},
                {J{\o}rgensen}, {Nordlund}, \& {Plez}}]{Gustafsson2008}
        {Gustafsson}, B., {Edvardsson}, B., {Eriksson}, K., {et~al.} 2008, \aap, 486,
        951
        
        \bibitem[{{Hasselquist} {et~al.}(2021){Hasselquist}, {Hayes}, {Lian},
                {Weinberg}, {Zasowski}, {Horta}, {Beaton}, {Feuillet}, {Garro}, {Gallart},
                {Smith}, {Holtzman}, {Minniti}, {Shetrone}, {J{\"o}nsson}, {Cioni},
                {Fillingham}, {Cunha}, {O{\'C}onnell}, {Fern{\'a}ndez-Trincado}, {Mu{\~n}oz},
                {Schiavon}, {Almeida}, {Anguiano}, {Beers}, {Bizyaev}, {Brownstein}, {Cohen},
                {Frinchaboy}, {Garc{\'\i}a-Hern{\'a}ndez}, {Geisler}, {Lane}, {Majewski},
                {Nidever}, {Nitschelm}, {Povick}, {Price-Whelan}, {Roman-Lopes}, {Rosado},
                {Sobeck}, {Stringfellow}, {Valenzuela}, {Villanova}, \&
                {Vincenzo}}]{Hasselquist2021}
        {Hasselquist}, S., {Hayes}, C.~R., {Lian}, J., {et~al.} 2021, arXiv e-prints,
        arXiv:2109.05130
        
        \bibitem[{{Hasselquist} {et~al.}(2017){Hasselquist}, {Shetrone}, {Smith},
                {Holtzman}, {McWilliam}, {Fern{\'a}ndez-Trincado}, {Beers}, {Majewski},
                {Nidever}, {Tang}, {Tissera}, {Fern{\'a}ndez Alvar}, {Allende Prieto},
                {Almeida}, {Anguiano}, {Battaglia}, {Carigi}, {Delgado Inglada},
                {Frinchaboy}, {Garc{\'\i}a-Hern{\'a}ndez}, {Geisler}, {Minniti}, {Placco},
                {Schultheis}, {Sobeck}, \& {Villanova}}]{Hasselquist2017}
        {Hasselquist}, S., {Shetrone}, M., {Smith}, V., {et~al.} 2017, \apj, 845, 162
        
        \bibitem[{{Holtzman} {et~al.}(2015){Holtzman}, {Shetrone}, {Johnson}, {Allende
                        Prieto}, {Anders}, {Andrews}, {Beers}, {Bizyaev}, {Blanton}, {Bovy},
                {Carrera}, {Chojnowski}, {Cunha}, {Eisenstein}, {Feuillet}, {Frinchaboy},
                {Galbraith-Frew}, {Garc{\'\i}a P{\'e}rez}, {Garc{\'\i}a-Hern{\'a}ndez},
                {Hasselquist}, {Hayden}, {Hearty}, {Ivans}, {Majewski}, {Martell},
                {Meszaros}, {Muna}, {Nidever}, {Nguyen}, {O'Connell}, {Pan}, {Pinsonneault},
                {Robin}, {Schiavon}, {Shane}, {Sobeck}, {Smith}, {Troup}, {Weinberg},
                {Wilson}, {Wood-Vasey}, {Zamora}, \& {Zasowski}}]{Holtzman2015}
        {Holtzman}, J.~A., {Shetrone}, M., {Johnson}, J.~A., {et~al.} 2015, \aj, 150,
        148
        
        \bibitem[{{Kunder} {et~al.}(2021){Kunder}, {Crabb}, {Debattista},
                {Koch-Hansen}, \& {Huhmann}}]{Kunder2021}
        {Kunder}, A., {Crabb}, R.~E., {Debattista}, V.~P., {Koch-Hansen}, A.~J., \&
        {Huhmann}, B.~M. 2021, \aj, 162, 86
        
        \bibitem[{{Kunder} \& {Butler}(2020)}]{Kunder2020}
        {Kunder}, A.~M. \& {Butler}, E. 2020, \aj, 160, 241
        
        \bibitem[{{Majewski} {et~al.}(2017){Majewski}, {Schiavon}, {Frinchaboy},
                {Allende Prieto}, {Barkhouser}, {Bizyaev}, {Blank}, {Brunner}, {Burton},
                {Carrera}, {Chojnowski}, {Cunha}, {Epstein}, {Fitzgerald}, {Garc{\'\i}a
                        P{\'e}rez}, {Hearty}, {Henderson}, {Holtzman}, {Johnson}, {Lam}, {Lawler},
                {Maseman}, {M{\'e}sz{\'a}ros}, {Nelson}, {Nguyen}, {Nidever}, {Pinsonneault},
                {Shetrone}, {Smee}, {Smith}, {Stolberg}, {Skrutskie}, {Walker}, {Wilson},
                {Zasowski}, {Anders}, {Basu}, {Beland}, {Blanton}, {Bovy}, {Brownstein},
                {Carlberg}, {Chaplin}, {Chiappini}, {Eisenstein}, {Elsworth}, {Feuillet},
                {Fleming}, {Galbraith-Frew}, {Garc{\'\i}a}, {Garc{\'\i}a-Hern{\'a}ndez},
                {Gillespie}, {Girardi}, {Gunn}, {Hasselquist}, {Hayden}, {Hekker}, {Ivans},
                {Kinemuchi}, {Klaene}, {Mahadevan}, {Mathur}, {Mosser}, {Muna}, {Munn},
                {Nichol}, {O'Connell}, {Parejko}, {Robin}, {Rocha-Pinto}, {Schultheis},
                {Serenelli}, {Shane}, {Silva Aguirre}, {Sobeck}, {Thompson}, {Troup},
                {Weinberg}, \& {Zamora}}]{Majewski2017}
        {Majewski}, S.~R., {Schiavon}, R.~P., {Frinchaboy}, P.~M., {et~al.} 2017, \aj,
        154, 94
        
        \bibitem[{{Masseron} {et~al.}(2016){Masseron}, {Merle}, \&
                {Hawkins}}]{Masseron2016}
        {Masseron}, T., {Merle}, T., \& {Hawkins}, K. 2016, {BACCHUS: Brussels
                Automatic Code for Characterizing High accUracy Spectra}
        
        \bibitem[{{M{\'e}sz{\'a}ros} {et~al.}(2020){M{\'e}sz{\'a}ros}, {Masseron},
                {Garc{\'\i}a-Hern{\'a}ndez}, {Allende Prieto}, {Beers}, {Bizyaev},
                {Chojnowski}, {Cohen}, {Cunha}, {Dell'Agli}, {Ebelke},
                {Fern{\'a}ndez-Trincado}, {Frinchaboy}, {Geisler}, {Hasselquist}, {Hearty},
                {Holtzman}, {Johnson}, {Lane}, {Lacerna}, {Longa-Pe{\~n}a}, {Majewski},
                {Martell}, {Minniti}, {Nataf}, {Nidever}, {Pan}, {Schiavon}, {Shetrone},
                {Smith}, {Sobeck}, {Stringfellow}, {Szigeti}, {Tang}, {Wilson}, \&
                {Zamora}}]{Meszaros2020}
        {M{\'e}sz{\'a}ros}, S., {Masseron}, T., {Garc{\'\i}a-Hern{\'a}ndez}, D.~A.,
        {et~al.} 2020, \mnras, 492, 1641
        
        \bibitem[{{Minniti}(2018)}]{Minniti2018a}
        {Minniti}, D. 2018, in The Vatican Observatory, Castel Gandolfo: 80th
        Anniversary Celebration, ed. G.~{Gionti} \& J.-B. {Kikwaya Eluo}, Vol.~51, 63
        
        \bibitem[{{Minniti} {et~al.}(2021{\natexlab{a}}){Minniti},
                {Fern{\'a}ndez-Trincado}, {G{\'o}mez}, {Smith}, {Lucas}, \& {Contreras
                        Ramos}}]{Minniti2021b}
        {Minniti}, D., {Fern{\'a}ndez-Trincado}, J.~G., {G{\'o}mez}, M., {et~al.}
        2021{\natexlab{a}}, \aap, 650, L11
        
        \bibitem[{{Minniti} {et~al.}(2018{\natexlab{a}}){Minniti},
                {Fern{\'a}ndez-Trincado}, {Ripepi}, {Alonso-Garc{\'\i}a}, {Contreras Ramos},
                \& {Marconi}}]{Minniti2018b}
        {Minniti}, D., {Fern{\'a}ndez-Trincado}, J.~G., {Ripepi}, V., {et~al.}
        2018{\natexlab{a}}, \apjl, 869, L10
        
        \bibitem[{{Minniti} {et~al.}(2021{\natexlab{b}}){Minniti},
                {Fern{\'a}ndez-Trincado}, {Smith}, {Lucas}, {G{\'o}mez}, \&
                {Pullen}}]{Minniti2021}
        {Minniti}, D., {Fern{\'a}ndez-Trincado}, J.~G., {Smith}, L.~C., {et~al.}
        2021{\natexlab{b}}, \aap, 648, A86
        
        \bibitem[{{Minniti} {et~al.}(2017{\natexlab{a}}){Minniti}, {Geisler},
                {Alonso-Garc{\'\i}a}, {Palma}, {Beam{\'\i}n}, {Borissova}, {Catelan},
                {Clari{\'a}}, {Cohen}, {Contreras Ramos}, {Dias}, {Fern{\'a}ndez-Trincado},
                {G{\'o}mez}, {Hempel}, {Ivanov}, {Kurtev}, {Lucas}, {Moni-Bidin}, {Pullen},
                {Ram{\'\i}rez Alegr{\'\i}a}, {Saito}, \& {Valenti}}]{Minniti2017c}
        {Minniti}, D., {Geisler}, D., {Alonso-Garc{\'\i}a}, J., {et~al.}
        2017{\natexlab{a}}, \apjl, 849, L24
        
        \bibitem[{{Minniti} {et~al.}(2020){Minniti}, {G{\'o}mez}, {Pullen}, {Palma},
                {Clari{\'a}}, {Alonso-Garc{\'\i}a}, {Saito}, {Smith},
                {Fern{\'a}ndez-Trincado}, \& {Hempel}}]{Minniti2020b}
        {Minniti}, D., {G{\'o}mez}, M., {Pullen}, J.~B., {et~al.} 2020, Research Notes
        of the American Astronomical Society, 4, 218
        
        \bibitem[{{Minniti} {et~al.}(2010){Minniti}, {Lucas}, {Emerson}, {Saito},
                {Hempel}, {Pietrukowicz}, {Ahumada}, {Alonso}, {Alonso-Garcia}, {Arias},
                {Bandyopadhyay}, {Barb{\'a}}, {Barbuy}, {Bedin}, {Bica}, {Borissova},
                {Bronfman}, {Carraro}, {Catelan}, {Clari{\'a}}, {Cross}, {de Grijs},
                {D{\'e}k{\'a}ny}, {Drew}, {Fari{\~n}a}, {Feinstein}, {Fern{\'a}ndez
                        Laj{\'u}s}, {Gamen}, {Geisler}, {Gieren}, {Goldman}, {Gonzalez}, {Gunthardt},
                {Gurovich}, {Hambly}, {Irwin}, {Ivanov}, {Jord{\'a}n}, {Kerins}, {Kinemuchi},
                {Kurtev}, {L{\'o}pez-Corredoira}, {Maccarone}, {Masetti}, {Merlo},
                {Messineo}, {Mirabel}, {Monaco}, {Morelli}, {Padilla}, {Palma}, {Parisi},
                {Pignata}, {Rejkuba}, {Roman-Lopes}, {Sale}, {Schreiber}, {Schr{\"o}der},
                {Smith}, {}, {Soto}, {Tamura}, {Tappert}, {Thompson}, {Toledo}, {Zoccali}, \&
                {Pietrzynski}}]{Minniti2010}
        {Minniti}, D., {Lucas}, P.~W., {Emerson}, J.~P., {et~al.} 2010, \na, 15, 433
        
        \bibitem[{{Minniti} {et~al.}(2017{\natexlab{b}}){Minniti}, {Palma},
                {D{\'e}k{\'a}ny}, {Hempel}, {Rejkuba}, {Pullen}, {Alonso-Garc{\'\i}a},
                {Barb{\'a}}, {Barbuy}, {Bica}, {Bonatto}, {Borissova}, {Catelan},
                {Carballo-Bello}, {Chene}, {Clari{\'a}}, {Cohen}, {Contreras Ramos}, {Dias},
                {Emerson}, {Froebrich}, {Buckner}, {Geisler}, {Gonzalez}, {Gran}, {Hajdu},
                {Irwin}, {Ivanov}, {Kurtev}, {Lucas}, {Majaess}, {Mauro}, {Moni-Bidin},
                {Navarrete}, {Ram{\'\i}rez Alegr{\'\i}a}, {Saito}, {Valenti}, \&
                {Zoccali}}]{Minniti2017d}
        {Minniti}, D., {Palma}, T., {D{\'e}k{\'a}ny}, I., {et~al.} 2017{\natexlab{b}},
        \apjl, 838, L14
        
        \bibitem[{{Minniti} {et~al.}(2021{\natexlab{c}}){Minniti}, {Ripepi},
                {Fern{\'a}ndez-Trincado}, {Alonso-Garc{\'\i}a}, {Smith}, {Lucas},
                {G{\'o}mez}, {Pullen}, {Garro}, {Vivanco C{\'a}diz}, {Hempel}, {Rejkuba},
                {Saito}, {Palma}, {Clari{\'a}}, {Gregg}, \& {Majaess}}]{Minniti2021c}
        {Minniti}, D., {Ripepi}, V., {Fern{\'a}ndez-Trincado}, J.~G., {et~al.}
        2021{\natexlab{c}}, \aap, 647, L4
        
        \bibitem[{{Minniti} {et~al.}(2018{\natexlab{b}}){Minniti}, {Schlafly}, {Palma},
                {Clari{\'a}}, {Hempel}, {Alonso-Garc{\'\i}a}, {Bica}, {Bonatto}, {Braga},
                {Clementini}, {Garofalo}, {G{\'o}mez}, {Ivanov}, {Lucas}, {Pullen}, {Saito},
                \& {Smith}}]{Minniti2018c}
        {Minniti}, D., {Schlafly}, E.~F., {Palma}, T., {et~al.} 2018{\natexlab{b}},
        \apj, 866, 12
        
        \bibitem[{{Nidever} {et~al.}(2015){Nidever}, {Holtzman}, {Allende Prieto},
                {Beland}, {Bender}, {Bizyaev}, {Burton}, {Desphande}, {Fleming}, {Garc{\'\i}a
                        P{\'e}rez}, {Hearty}, {Majewski}, {M{\'e}sz{\'a}ros}, {Muna}, {Nguyen},
                {Schiavon}, {Shetrone}, {Skrutskie}, {Sobeck}, \& {Wilson}}]{Nidever2015}
        {Nidever}, D.~L., {Holtzman}, J.~A., {Allende Prieto}, C., {et~al.} 2015, \aj,
        150, 173
        
        \bibitem[{{Obasi} {et~al.}(2021){Obasi}, {G{\'o}mez}, {Minniti}, \&
                {Alonso-Garc{\'\i}a}}]{Obasi2021}
        {Obasi}, C., {G{\'o}mez}, M., {Minniti}, D., \& {Alonso-Garc{\'\i}a}, J. 2021,
        \aap, 654, A39
        
        \bibitem[{{O'Donnell}(1994)}]{Donnell1994}
        {O'Donnell}, J.~E. 1994, \apj, 422, 158
        
        \bibitem[{{Palma} {et~al.}(2019){Palma}, {Minniti}, {Alonso-Garc{\'\i}a},
                {Crestani}, {Netzel}, {Clari{\'a}}, {Saito}, {Dias},
                {Fern{\'a}ndez-Trincado}, {Kammers}, {Geisler}, {G{\'o}mez}, {Hempel}, \&
                {Pullen}}]{Palma2019}
        {Palma}, T., {Minniti}, D., {Alonso-Garc{\'\i}a}, J., {et~al.} 2019, \mnras,
        487, 3140
        
        \bibitem[{{P{\'e}rez-Villegas} {et~al.}(2020){P{\'e}rez-Villegas}, {Barbuy},
                {Kerber}, {Ortolani}, {Souza}, \& {Bica}}]{Perez-Villegas2020}
        {P{\'e}rez-Villegas}, A., {Barbuy}, B., {Kerber}, L.~O., {et~al.} 2020, \mnras,
        491, 3251
        
        \bibitem[{{Romero-Colmenares} {et~al.}(2021){Romero-Colmenares},
                {Fern{\'a}ndez-Trincado}, {Geisler}, {Souza}, {Villanova}, {Longa-Pe{\~n}a},
                {Minniti}, {Beers}, {Bidin}, {Perez-Villegas}, {Moreno}, {Garro}, {Baeza},
                {Henao}, {Barbuy}, {Alonso-Garc{\'\i}a}, {Cohen}, {Lane}, \&
                {Mu{\~n}oz}}]{Romero-Colmenares2021}
        {Romero-Colmenares}, M., {Fern{\'a}ndez-Trincado}, J.~G., {Geisler}, D.,
        {et~al.} 2021, \aap, 652, A158
        
        \bibitem[{{Saito} {et~al.}(2012){Saito}, {Hempel}, {Minniti}, {Lucas},
                {Rejkuba}, {Toledo}, {Gonzalez}, {Alonso-Garc{\'\i}a}, {Irwin},
                {Gonzalez-Solares}, {Hodgkin}, {Lewis}, {Cross}, {Ivanov}, {Kerins},
                {Emerson}, {Soto}, {Am{\^o}res}, {Gurovich}, {D{\'e}k{\'a}ny}, {Angeloni},
                {Beamin}, {Catelan}, {Padilla}, {Zoccali}, {Pietrukowicz}, {Moni Bidin},
                {Mauro}, {Geisler}, {Folkes}, {Sale}, {Borissova}, {Kurtev}, {Ahumada},
                {Alonso}, {Adamson}, {Arias}, {Bandyopadhyay}, {Barb{\'a}}, {Barbuy},
                {Baume}, {Bedin}, {Bellini}, {Benjamin}, {Bica}, {Bonatto}, {Bronfman},
                {Carraro}, {Chen{\`e}}, {Clari{\'a}}, {Clarke}, {Contreras}, {Corvill{\'o}n},
                {de Grijs}, {Dias}, {Drew}, {Fari{\~n}a}, {Feinstein},
                {Fern{\'a}ndez-Laj{\'u}s}, {Gamen}, {Gieren}, {Goldman},
                {Gonz{\'a}lez-Fern{\'a}ndez}, {Grand}, {Gunthardt}, {Hambly}, {Hanson},
                {He{\l}miniak}, {Hoare}, {Huckvale}, {Jord{\'a}n}, {Kinemuchi}, {Longmore},
                {L{\'o}pez-Corredoira}, {Maccarone}, {Majaess}, {Mart{\'\i}n}, {Masetti},
                {Mennickent}, {Mirabel}, {Monaco}, {Morelli}, {Motta}, {Palma}, {Parisi},
                {Parker}, {Pe{\~n}aloza}, {Pietrzy{\'n}ski}, {Pignata}, {Popescu}, {Read},
                {Rojas}, {Roman-Lopes}, {Ruiz}, {Saviane}, {Schreiber}, {Schr{\"o}der},
                {Sharma}, {Smith}, {Sodr{\'e}}, {Stead}, {Stephens}, {Tamura}, {Tappert},
                {Thompson}, {Valenti}, {Vanzi}, {Walton}, {Weidmann}, \&
                {Zijlstra}}]{Saito2012}
        {Saito}, R.~K., {Hempel}, M., {Minniti}, D., {et~al.} 2012, \aap, 537, A107
        
        \bibitem[{{Sanders} {et~al.}(2019){Sanders}, {Smith}, \& {Evans}}]{Sanders2019}
        {Sanders}, J.~L., {Smith}, L., \& {Evans}, N.~W. 2019, \mnras, 488, 4552
        
        \bibitem[{{Santana} {et~al.}(2021){Santana}, {Beaton}, {Covey}, {O'Connell},
                {Longa-Pe{\~n}a}, {Cohen}, {Fern{\'a}ndez-Trincado}, {Hayes}, {Zasowski},
                {Sobeck}, {Majewski}, {Chojnowski}, {De Lee}, {Oelkers}, {Stringfellow},
                {Almeida}, {Anguiano}, {Donor}, {Frinchaboy}, {Hasselquist}, {Johnson},
                {Kollmeier}, {Nidever}, {Price-Whelan}, {Rojas-Arriagada}, {Schultheis},
                {Shetrone}, {Simon}, {Aerts}, {Borissova}, {Drout}, {Geisler}, {Law},
                {Medina}, {Minniti}, {Monachesi}, {Mu{\~n}oz}, {Poleski}, {Roman-Lopes},
                {Schlaufman}, {Stutz}, {Teske}, {Tkachenko}, {Van Saders}, {Weinberger}, \&
                {Zoccali}}]{Santana2021}
        {Santana}, F.~A., {Beaton}, R.~L., {Covey}, K.~R., {et~al.} 2021, arXiv
        e-prints, arXiv:2108.11908
        
        \bibitem[{{Saracino} {et~al.}(2015){Saracino}, {Dalessandro}, {Ferraro},
                {Lanzoni}, {Geisler}, {Mauro}, {Villanova}, {Moni Bidin}, {Miocchi}, \&
                {Massari}}]{Saracino2015}
        {Saracino}, S., {Dalessandro}, E., {Ferraro}, F.~R., {et~al.} 2015, \apj, 806,
        152
        
        \bibitem[{{Schiavon} {et~al.}(2017){Schiavon}, {Johnson}, {Frinchaboy},
                {Zasowski}, {M{\'e}sz{\'a}ros}, {Garc{\'\i}a-Hern{\'a}ndez}, {Cohen}, {Tang},
                {Villanova}, {Geisler}, {Beers}, {Fern{\'a}ndez-Trincado}, {Garc{\'\i}a
                        P{\'e}rez}, {Lucatello}, {Majewski}, {Martell}, {O'Connell}, {Allende
                        Prieto}, {Bizyaev}, {Carrera}, {Lane}, {Malanushenko}, {Malanushenko},
                {Mu{\~n}oz}, {Nitschelm}, {Oravetz}, {Pan}, {Roman-Lopes}, {Schultheis}, \&
                {Simmons}}]{Schiavon2017}
        {Schiavon}, R.~P., {Johnson}, J.~A., {Frinchaboy}, P.~M., {et~al.} 2017,
        \mnras, 466, 1010
        
        \bibitem[{{Schlafly} \& {Finkbeiner}(2011)}]{Schlafly2011}
        {Schlafly}, E.~F. \& {Finkbeiner}, D.~P. 2011, \apj, 737, 103
        
        \bibitem[{{Shetrone} {et~al.}(2001){Shetrone}, {C{\^o}t{\'e}}, \&
                {Sargent}}]{Shetrone2001}
        {Shetrone}, M.~D., {C{\^o}t{\'e}}, P., \& {Sargent}, W.~L.~W. 2001, \apj, 548,
        592
        
        \bibitem[{{Skrutskie} {et~al.}(2006){Skrutskie}, {Cutri}, {Stiening},
                {Weinberg}, {Schneider}, {Carpenter}, {Beichman}, {Capps}, {Chester},
                {Elias}, {Huchra}, {Liebert}, {Lonsdale}, {Monet}, {Price}, {Seitzer},
                {Jarrett}, {Kirkpatrick}, {Gizis}, {Howard}, {Evans}, {Fowler}, {Fullmer},
                {Hurt}, {Light}, {Kopan}, {Marsh}, {McCallon}, {Tam}, {Van Dyk}, \&
                {Wheelock}}]{Skrutskie2006}
        {Skrutskie}, M.~F., {Cutri}, R.~M., {Stiening}, R., {et~al.} 2006, \aj, 131,
        1163
        
        \bibitem[{{Smith} {et~al.}(2021){Smith}, {Bizyaev}, {Cunha}, {Shetrone},
                {Souto}, {Allende Prieto}, {Masseron}, {M{\'e}sz{\'a}ros}, {J{\"o}nsson},
                {Hasselquist}, {Osorio}, {Garc{\'\i}a-Hern{\'a}ndez}, {Plez}, {Beaton},
                {Holtzman}, {Majewski}, {Stringfellow}, \& {Sobeck}}]{Smith2021}
        {Smith}, V.~V., {Bizyaev}, D., {Cunha}, K., {et~al.} 2021, \aj, 161, 254
        
        \bibitem[{{Sofue}(2015)}]{Sofue2015}
        {Sofue}, Y. 2015, \pasj, 67, 75
        
        \bibitem[{{Villanova} {et~al.}(2019){Villanova}, {Monaco}, {Geisler},
                {O'Connell}, {Minniti}, {Assmann}, \& {Barb{\'a}}}]{Villanova2019}
        {Villanova}, S., {Monaco}, L., {Geisler}, D., {et~al.} 2019, \apj, 882, 174
        
        \bibitem[{{Wilson} {et~al.}(2019){Wilson}, {Hearty}, {Skrutskie}, {Majewski},
                {Holtzman}, {Eisenstein}, {Gunn}, {Blank}, {Henderson}, {Smee}, {Nelson},
                {Nidever}, {Arns}, {Barkhouser}, {Barr}, {Beland}, {Bershady}, {Blanton},
                {Brunner}, {Burton}, {Carey}, {Carr}, {Colque}, {Crane}, {Damke}, {Davidson},
                {Dean}, {Di Mille}, {Don}, {Ebelke}, {Evans}, {Fitzgerald}, {Gillespie},
                {Hall}, {Harding}, {Harding}, {Hammond}, {Hancock}, {Harrison}, {Hope},
                {Horne}, {Karakla}, {Lam}, {Leger}, {MacDonald}, {Maseman}, {Matsunari},
                {Melton}, {Mitcheltree}, {O'Brien}, {O'Connell}, {Patten}, {Richardson},
                {Rieke}, {Rieke}, {Roman-Lopes}, {Schiavon}, {Sobeck}, {Stolberg}, {Stoll},
                {Tembe}, {Trujillo}, {Uomoto}, {Vernieri}, {Walker}, {Weinberg}, {Young},
                {Anthony-Brumfield}, {Bizyaev}, {Breslauer}, {De Lee}, {Downey}, {Halverson},
                {Huehnerhoff}, {Klaene}, {Leon}, {Long}, {Mahadevan}, {Malanushenko},
                {Nguyen}, {Owen}, {S{\'a}nchez-Gallego}, {Sayres}, {Shane}, {Shectman},
                {Shetrone}, {Skinner}, {Stauffer}, \& {Zhao}}]{Wilson2019}
        {Wilson}, J.~C., {Hearty}, F.~R., {Skrutskie}, M.~F., {et~al.} 2019, \pasp,
        131, 055001
        
        \bibitem[{{Zamora} {et~al.}(2015){Zamora}, {Garc{\'\i}a-Hern{\'a}ndez},
                {Allende Prieto}, {Carrera}, {Koesterke}, {Edvardsson}, {Castelli}, {Plez},
                {Bizyaev}, {Cunha}, {Garc{\'\i}a P{\'e}rez}, {Gustafsson}, {Holtzman},
                {Lawler}, {Majewski}, {Manchado}, {M{\'e}sz{\'a}ros}, {Shane}, {Shetrone},
                {Smith}, \& {Zasowski}}]{Zamora2015}
        {Zamora}, O., {Garc{\'\i}a-Hern{\'a}ndez}, D.~A., {Allende Prieto}, C.,
        {et~al.} 2015, \aj, 149, 181
        
        \bibitem[{{Zasowski} {et~al.}(2017){Zasowski}, {Cohen}, {Chojnowski},
                {Santana}, {Oelkers}, {Andrews}, {Beaton}, {Bender}, {Bird}, {Bovy},
                {Carlberg}, {Covey}, {Cunha}, {Dell'Agli}, {Fleming}, {Frinchaboy},
                {Garc{\'\i}a-Hern{\'a}ndez}, {Harding}, {Holtzman}, {Johnson}, {Kollmeier},
                {Majewski}, {M{\'e}sz{\'a}ros}, {Munn}, {Mu{\~n}oz}, {Ness}, {Nidever},
                {Poleski}, {Rom{\'a}n-Z{\'u}{\~n}iga}, {Shetrone}, {Simon}, {Smith},
                {Sobeck}, {Stringfellow}, {Szigeti{\'a}ros}, {Tayar}, \&
                {Troup}}]{Zasowski2017}
        {Zasowski}, G., {Cohen}, R.~E., {Chojnowski}, S.~D., {et~al.} 2017, \aj, 154,
        198
        
\end{thebibliography}

\end{document}